\documentclass[review]{elsarticle}

\usepackage{amssymb,amsmath}
\usepackage{extarrows}
\usepackage{float}
\usepackage{stix}
\usepackage{graphicx}
\usepackage{array}
\usepackage{amsbsy}
\usepackage[dvips]{epsfig}
\usepackage{bbm}
\usepackage{hhline}
\usepackage{multirow}
\usepackage{array}
\usepackage[colorlinks,linkcolor=blue,citecolor=blue,urlcolor=blue]{hyperref}

\renewcommand{\[}{\left[}
\renewcommand{\]}{\right]}
\renewcommand{\(}{\left(}
\renewcommand{\)}{\right)}
\def\nl         {\right.\\ \left.}
\def\lb         {\left|}
\def\rb         {\right|}

\def\rar        {\rightarrow}  

\def\Rar        {\Rightarrow}

\newcommand{\eq}[1]{\begin{align}#1\end{align}}
\newcommand{\ml}[1]{\begin{multline}#1\end{multline}}
\newcommand{\eqg}[1]{\begin{gather}#1\end{gather}}
\newcommand{\seq}[1]{\begin{subequations}#1\end{subequations}}

\DeclareMathOperator{\sign}{sign}

\newcommand{\stkout}[1]{\ifmmode\text{\sout{\ensuremath{#1}}}\else\sout{#1}\fi}

\newcommand{\bra}[1]{\mbox{$\langle #1 |$}}
\newcommand{\ket}[1]{\mbox{$| #1 \rangle$}}
\newcommand{\average}[1]{\left\langle #1 \right\rangle}

\newcommand{\lab}[1]{\label{#1}}

\newcommand{\uuu}[1]{\underline{\underline{#1}}}
\newcommand{\uu}[1]{\underline{#1}}
\newcommand{\oo}[1]{\overline{#1}}
\newcommand{\seclab}[1]{Sec.~\ref{#1}}
\newcommand{\e}[1]{Eq.~\eqref{#1}}
\newcommand{\erange}[2]{Eq.~\eqref{#1}--\eqref{#2}}
\newcommand{\elab}[2]{Eq.(\ref{#1}#2)}
\newcommand{\fig}[1]{Fig.\ref{#1}}
\newcommand{\figlab}[2]{Fig.\ref{#1}(#2)}
\renewcommand{\sec}[1]{Section\,\ref{#1}}

\newcommand{\h}[1]{\hat{#1}}

\newcommand{\mr}[1]{{\mathrm #1}}
\newcommand{\ocite}[1]{Ref.\cite{#1}}
\def\Re{{\rm Re}}
\def\Im{{\rm Im}}
\renewcommand{\d}{\mr{d}}           
\def\ppe        {P\&p\ }
\def\ai	        {{\em ab--initio}}

\def\yambo      {{\it Yambo}~\cite{Sangalli_2019,AndreaMarini2009}}

\def\gc         {\gamma}

\def\gd         {\delta}

\def\gee        {\epsilon}

\def\go         {\omega}

\def\gr         {\rho}

\def\gS         {\Sigma}

\def\gt         {\theta}

\def\pp		{{\mathbf p}}

\def\rr		{{\mathbf r}}

\def\kk		{{\mathbf k}}
\def\qq		{{\mathbf q}}

\def\D{\Delta}

\def\h{\eta}
\def\th{\theta}
\def\m{\mu}

\def\p{\pi}
\def\r{\rho}

\def\s{\sigma}
\def\t{T}
\def\vf{\varphi}
\def\x{\chi}
\def\w{\omega}

\def\blj{{\mathbf j}}
\def\blk{{\mathbf k}}

\def\blp{{\mathbf p}}

\def\blr{{\mathbf r}}

\def\blv{{\mathbf v}}

\def\blA{{\mathbf A}}

\def\blE{{\mathbf E}}

\def\callL{\mbox{$\mathcal{L}$}}

\newcommand{\cnrism} {Istituto di Struttura della Materia and Division of Ultrafast Processes in Materials (FLASHit) of the National Research Council, via Salaria Km 29.3, I-00016 Monterotondo Stazione, Italy}
\newcommand{\etsf} {European Theoretical Spectroscopy Facilities (ETSF)}
\newcommand{\tov} {Dipartimento di Fisica, Universit\`{a} di Roma Tor Vergata, Via della Ricerca Scientifica 1, 00133 Rome, Italy}

\begin{document}

\begin{frontmatter}
\title{Coherence and de--coherence in the Time--Resolved ARPES of realistic materials: an ab-initio perspective}

\author{Andrea Marini}
\address{\cnrism}
\address{\etsf}
\author{Enrico Perfetto}
\address{\tov}
\author{Gianluca Stefanucci}
\address{\tov}

\begin{abstract}
Coherence and de--coherence are the most fundamental steps that follow the initial photo--excitation occurring in typical \ppe experiments. Indeed, the initial
external laser pulse transfers coherence to the system in terms of creation of multiple electron--hole pairs excitation. The excitation concurs both to the
creation of a finite carriers density and to the appearance of induced electromagnetic fields. The two effects, to a very first approximation, can be connected
to the simple concepts of populations and oscillations. The dynamics of the system following the initial photo--excitation is, thus, entirely dictated by the
interplay between coherence and de--coherence. This interplay and the de--coherence process itself, is due to the correlation effects stimulated by the
photo--excitation. Single--particle, like the electron--phonon, and two--particles, like the electron--electron, scattering processes induce a complex dynamics
of the electrons that, in turn, makes the description of the correlated and photo--excited system in terms of pure excitonic and/or carriers populations
challenging. 
\end{abstract}
\end{frontmatter}

\tableofcontents

\section{Introduction}
Time-resolved (TR) and angle-resolved photo-emission spectroscopy (ARPES) has established as a powerful experimental technique to monitor the ultrafast dynamics
of electronic excitation in materials. Applications cover  image potential states~\cite{GHHRS.1995,FW.1995,EBCFGH.2004,VMR.2004,CWAGGP.2014,SLD.2015} electron
relaxation in metals~\cite{FSTB.1992,PO.1997,SAEMMCG.1994,LLBSGW.2004} semiconductors~\cite{WKFR.2004,SS.2009,Netal.2014,WZR.2015} and more recently topological
insulators~\cite{RGKCH.2014,SYACFKS.2012,WHSSGLJG.2012,CRCZGBBKGP.2012,NOHFMEAABEC.2014,BVLNL.2014},  charge transfer processes at solid state
interfaces~\cite{GWLMGH.1998,VZ.1999,MYTZ.2008,ZYM.2009,VMT.2011} and in adsorbate on
surfaces~\cite{HBSW.1997,Setal.2002,ZGW.2002,OLP.2004,MSMYLBAMK.2008,TY.2005,Aetal.2003} and the formation and dynamics of
excitons~\cite{WKFR.2004,SS.2009,Z.2015,SRSW.1980,VBBT.2009,DWMRWS.2014}.

In TR--ARPES  on semiconductors or insulators a coherent laser pulse excites electrons from the valence band to the conduction band.  During pumping and for
about a few hundreds of femtoseconds after several semiconductors with a gap of 1 eV or larger~\cite{SM.2002,Betal.1995,BC.1997,sm.2015} coherently oscillate between the ground state and the
dipole-allowed excited states.
Due to the Coulomb attraction between the conduction electrons and the valence holes the excited states may also contain bound electron-hole
({\em eh}) pairs or {\em excitons}. These excitons are said {\em coherent} or {\em virtual}~\cite{HK-book.1993,SW-book.2002,KKKG.2006} since they give rise to
an oscillating (in time) polarization of the medium.  After tens of picoseconds coherence is destroyed by electron-electron~\cite{HaugKochbook,Steinhoff2016} and
electron-phonon~\cite{Selig2016,MolinaSanchez2017} scattering processes. The system reaches a quasi-equilibrium state characterized by quasi-free carriers
coexisting with  {\it incoherent} or {\rm real} excitons~\cite{Ulbricht2011,Koch2006,PSMS.2016}.  In this steady regime the electronic density matrix is not a
pure state but an admixture of the ground state and dressed electron-hole excited states.

The photocurrent of a TR--ARPES  experiment is generated by a  {\em probe} pulse impinging the system. The spectrum depends on the intensity and frequency of the
pump excitation, on the delay between pump and probe as well as on the duration of the probe~\cite{Perfetto2020a}. Therefore TR--ARPES provides a unique tool to
reveal different excitation and in particular to characterize the nature of what is generally defined an ``exciton''.  In fact, excitonic structures inside the
gap can change their shape, position and intensity, meaning that excited states with different excitonic character are visited.

In this work we will first give, in \sec{sec:motivations}, some motivations to the device of a microscopic theory of Many--Body systems at and out--of
equilibrium. We will discuss how a \ppe\, experiment can actually trigger phenomena specific to the excitation process and also induce the coherent formation of
high energy states. We will then introduce, in \sec{sec:H} the main mathematical ingredients of the Hamiltonian considered in this work. Two different reference
systems will be used: a fully \ai\, scheme based on Density--Functional Theory, \sec{sec:AI_connection}, and a two--bands model Hamiltonian, \sec{sec:model_H}, that will allow to
highlight some key aspects in a transparent way.

After reviewing, in
\sec{sec:MBPT} the equilibrium Many--Body theory in order to introduce key concepts like quasiparticles and excitons, we will move out--of--equilibrium in
\sec{sec:NEGF}. The different self--energies used both at equilibrium and out--of--equilibrium are introduced in \sec{sec:SEs}.

The third part of this work is entirely devoted to the TR--ARPES. We will first introduce the concepts of coherent and in--coherent regime in \sec{sec:coh}. We
will then use two very different approaches in the two regimes. We will first investigate coherent excitonic effects by using a simple model Hamiltonian in
\sec{sec:coh_excitons}. The same Hamiltonian will be used in \sec{sec:incoh_excitons} to describe the very different nature of bound electron--hole pairs in the
in--coherent regime, i.e. after the induced polarization has died. 

We will end this section with the \ai\, approach to TR--ARPES. The \ai\, NEGF approach will be first reviewed in \sec{sec:incoh_carriers} and later applied
to two paradigmatic materials: bulk Silicon (\sec{sec:Si}) and Black--Phosphorus (\sec{sec:BP}). 

\section{Motivations}
\lab{sec:motivations}
It would be very much desiderable to be able to interpret the photo--excited materials in terms of simple arguments, taken from the equilibrium physics. This,
indeed, is often based on simple concepts that can make the interpretation of the complex real--time dynamics appealing. As a matter of fact, instead, the
photo--excitation takes the system in am highly intricated state that, even by using advanced Many--Body concepts, is difficult to describe accurately.

This motivates a careful and deep investigation of the physics out--of--equilibrium especially when applied to realistic materials. In order to motivate this,
by using simple arguments, in the next two sections we will investigate two paradigmatic properties of a photo--excited material: the connection between
photo--excitation and coherence and the actual modification of the microscpic, elemental, scattering processes induced by the experimental pump field.

\subsection{Photo--induced coherence: preliminary remarks}
\begin{figure}[t]
\centering
\includegraphics[width=0.8\textwidth]{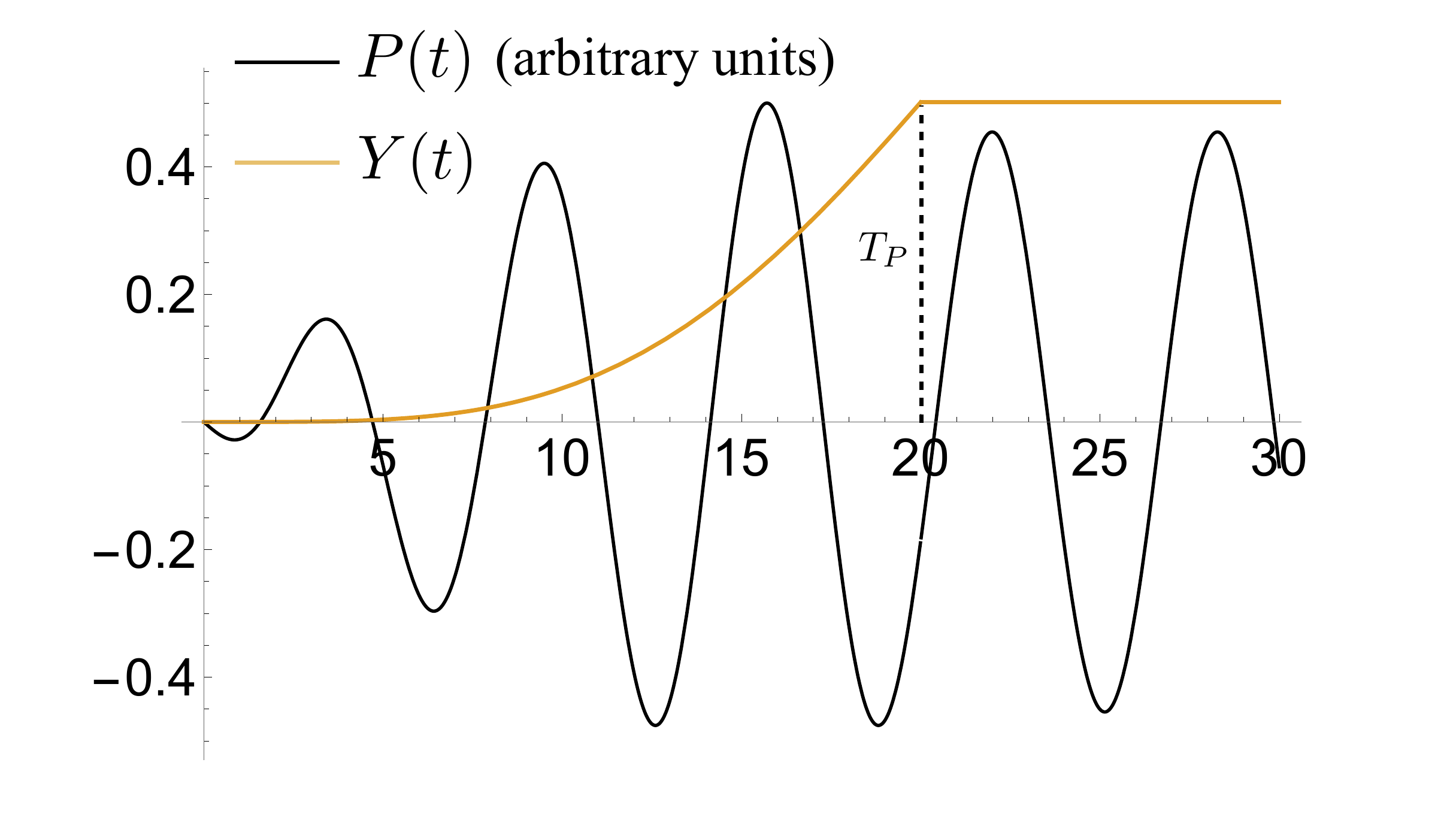}
\caption{Time--dependent  electron--hole pairs population, $Y\(t\)$ and polarization $P(t)$ in the simple model system described by \e{eq:coh.1}. This simple
model clearly show that the initial photo--excitation populates the empty bands and, at the same time, induces an oscillating contribution to the total electric
field.}
\label{coherencefig}
\end{figure}

To interpret ARPES spectra of photoexcited materials it is crucial to have a proper description of the nonequilibrium state of the system.  Depending on the
delay between the pump and probe pulses the nonequilibrium state of interest could be the one describing the system subject to an electromagnetic perturbation
(overlapping pump and probe), or shortly after the pump (coherent regime) or after the complete loss of coherence (incoherent regime). The differences between
these different regimes can already be highlighted using a simple two-level system, the lowest (highest) level representing the valence (conduction) band,
coupled through a dipole moment $D$ to an electric field $E\(t\)$:
\seq{
\lab{eq:coh.1}
\eqg{
\hat{H}(t)=\hat{h}_e+\hat{V}_{e-\gc}(t),\\
\hat{h}_e=\sum_{i=c,v}\gee_i \hat{d}^\dag_i\hat{d}_i ,\\
\hat{V}_{e-\gc}(t)=E(t)\Big(D\hat{d}^\dag_c\hat{d}_v+D^{\ast}\hat{d}^\dag_v\hat{d}_c\Big).
}}
Letting 
$|\Psi(t)\rangle=\Big(u_{c}(t)\hat{d}^{\dag}_{c}+u_{v}(t)\hat{d}^{\dag}_{v}\Big)|0\rangle$ 
be the state of the system at time $t$ 
the electronic density matrix reads
\begin{align}
\r_{ij}(t)=\langle \hat{d}^{\dag}_{j}(t)\hat{d}_{i}(t)\rangle=u^{\ast}_{j}(t)u_{i}(t),
\end{align}
whereas the population of electron-hole pairs~\cite{K.2000} reads
\eq{
Y\(t\)=\langle \hat{d}^{\dag}_{c}(t)\hat{d}_{v}(t) 
\hat{d}^{\dag}_{v}(t)\hat{d}_{c}(t)\rangle-\r_{vc}(t)\r_{cv}(t)=\r_{cc}(t)(1-\r_{vv}(t)).
}
For an initially filled valence band and a coherent pump field $E\(t\)=E_{0}\sin\w_{0}t$ the time-dependent amplitudes $u_{j}\(t\)$ can easily be worked out in
the rotating wave approximation~\cite{Leeuwen2013} and they read
\seq{
\eqg{
u_{c}(t)=-\frac{2iE_{0}D}{\sqrt{\Omega^{2}+4E_{0}^{2}|D|^{2}}}
e^{-i\Omega t/2}e^{-i\gee_{c}t}\sin\left(
\frac{\sqrt{\Omega^{2}+4E_{0}^{2}|D|^{2}}}{2}t\right),
\\
u_{v}(t)=ie^{i\Omega t}e^{-i\gee_{v}t}\frac{d}{dt}e^{i\gee_{c}t}\frac{u_{c}(t)}{E_{0}D},
}}
where we have defined $\Omega=\w_{0}-\gee_{c}+\gee_{v}$.  If the pump is switched off at time $T_{P}$ then for times $t>T_{P}$ we simply have
$u_{j}(t)=u_{j}(T_{P})e^{-i\gee_{j}(t-T_{P})}$.  In Fig.~\ref{coherencefig} we show the polarization of the system $P(t)= D\r_{cv}(t)+D^{\ast}\r_{vc}(t)$ as
well as the population of electron-hole pairs $Y(t)$ for $\gee_{c}-\gee_{v}=E_{g}$, $DE_{0}=0.1E_{g}$, $\Omega=0$ and $T_{P}=20$; energies are in units of the
gap $E_{g}$ and times are in units of $1/E_{g}$.

During pumping the laser transfers coherence to the system and after pumping, i.e., for times $t>T_{P}$, the polarization oscillates coherently. Although
$Y\(t\)$ attains a constant value the system is not in a quasi-thermal nor in a quasi-stationary state. This is the so called {\em coherent} regime and it can
last for up to few hundreds of femtoseconds.  Decoherence mechanisms typically due to electron-phonon scattering are responsible for a damping of the
polarization which eventually vanishes. This is the {\em incoherent} regime; the system can be described by a quasi-stationary density matrix. For a more
detailed  discussion see Ref.~\cite{Perfetto2015}.

\subsection{The paradigmatic case of the carrier lifetimes.  Are they really intrinsic quantities?}
\lab{sec:NEQ_lifetimes}
One of the common belief in time--resolved experiments is that the pump pulse produces an {\em ideal excited state}. What is this? It is an intrinsic 
excited state of system, that is independent on the excitation field. 

This aspect is very relevant to the physics of the material and to its potential application as technological device. This can be understood with a simple
example. Imagine you need to device a system where the carrier relaxation is super--fast. Then you use TR--ARPES to detect the relaxation time of carriers
pumped in the conduction bands. One paradigmatic example is described in \ocite{Ichibayashi2009} where, by using Time--Resolved two--photon
Photoemission\,(2PPE) it was deduced an ultrafast relaxation time $\tau_f\sim 40$\,fs of the photo--induced carriers.

The authors, however, interpreted this fast decay as due to intrinsic scattering mechanisms of the material, peculiar of the specific excited state triggered by
the laser pulse. Theoretically this corresponds to use the so called  Relaxation Time Approximation\cite{Bernardi2014} and reads:
\eq{
 \frac{d}{dt}f_i\(t\)=\gc^{EQ}_i\[f_i\(t_{th}\)-f_i\(t\)\],
 \lab{eq:02.03.1}
}
with $f_i\(t\)$ the occupation of the level $i$, $t_{th}$ the thermalization time where the occupations are supposed to follow a Fermi distribution with a given
temperature, and $\gc_i^{EQ}$ the {\em equilibrium} lifetimes computed with the presence of photo--excited carriers.

\e{eq:02.03.1} is often used as it allows to calculate the lifetimes once for all. This is a giant simplification that speeds up enormously the simulations.
At the same time \e{eq:02.03.1} conveys the message that the out--of--equilibrium dynamics is dictated by equilibrium quantities with the pump laser pulse just
promoting carriers in the conduction bands.

This is not true at all, at it was demonstrated in \ocite{Sangalli2015}. In this work the authors shows that the full carriers
equation of motion can be reduced to a non--linear equation for the carriers occupations that, however, differ from \e{eq:02.03.1}:
\eq{
 \frac{d}{dt}f_i\(t\)= \left. \frac{d}{dt}f_i\(t\)\right|_{pump}+\left.\frac{d}{dt}f_i\(t\)\right|_{relax}.
 \lab{eq:02.03.2}
}
In \e{eq:02.03.2} $\left. \frac{d}{dt}f_i\(t\)\right|_{pump}$ is the driving term, written in terms of the pump laser pulse. $\left.
\frac{d}{dt}f_i\(t\)\right|_{relax}$ is instead the scattering term, induced, for example by the scattering with the other electrons and/or phonons.

Therefore, at difference the RTA, \e{eq:02.03.2} describes also the excitation process that, in \e{eq:02.03.2} is replaced with an {\it ad--hoc} ansatz on the
photo--excited initial populations. Indeed from \ocite{Sangalli2015,Marini2013,Melo2016} we know that
\eq{
 \left. \frac{d}{dt}f_i\(t\)\right|_{relax}=-\gc^{\(e\)}_i\(t\)f_i\(t\)+\gc^{\(h\)}_i\(t\)\(1-f_i\(t\)\),
 \lab{eq:02.03.3}
}
with $\gc^{\(e/h\)}$ representing the electron/hole NEQ lifetimes. A crucial aspect of \e{eq:02.03.3} is the occupations and lifetimes are
time--dependent. It is natural to ask, then, to what extent $\gc^{\(e/h\)}_i\(t\)$ are different from their equilibrium counterpart. 

\begin{figure}[t]
\centering
\includegraphics[width=0.8\textwidth]{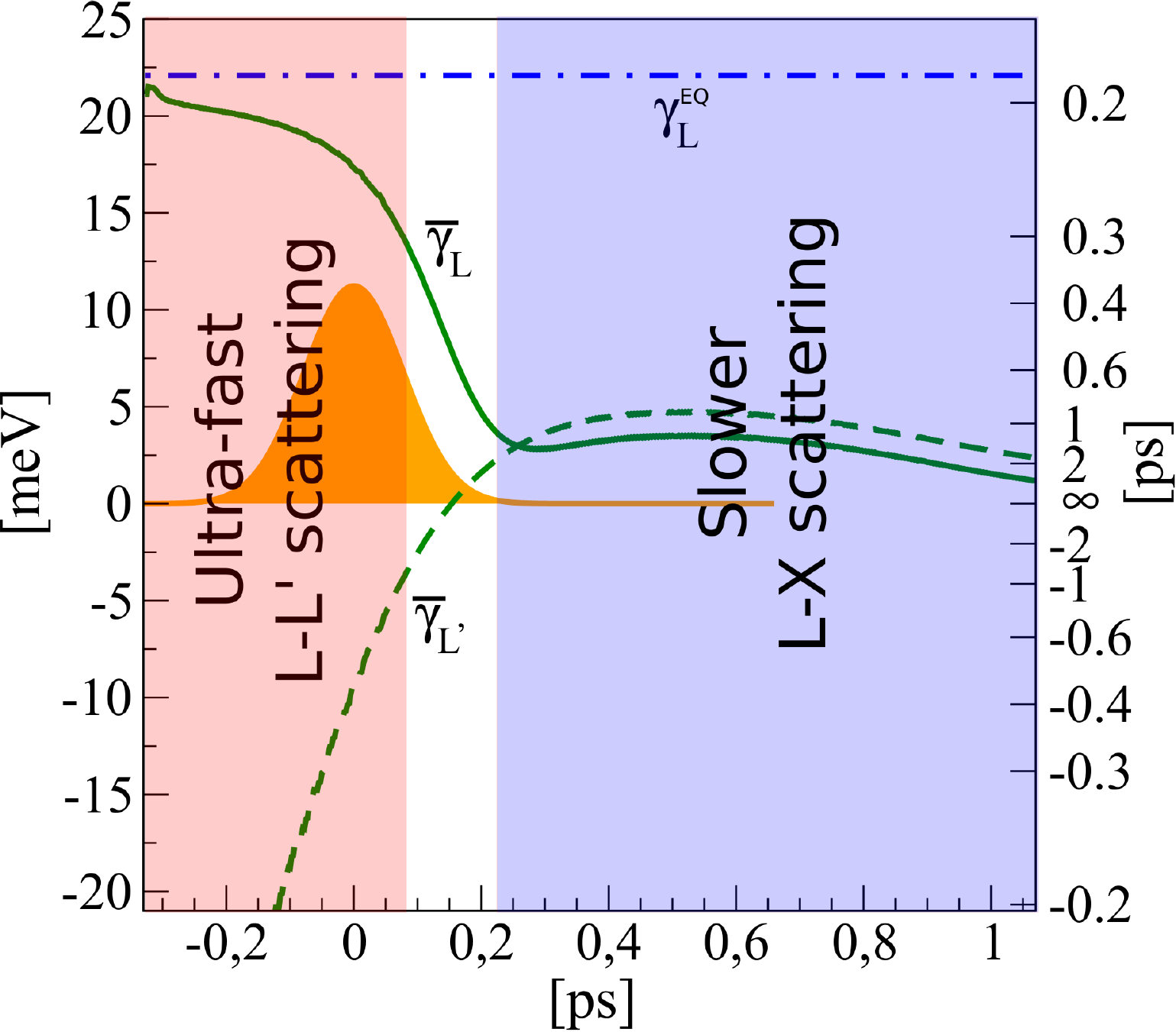}
\caption{Time--dependent carrier lifetimes ($\oo{\gc}_{L/L'}\(t\)$) compared with the equilibrium value ($\gc^{EQ}_{L}$) in bulk Silicon. From
\ocite{Sangalli2015}.}
\label{fig:02.03.1}
\end{figure}
In \fig{fig:02.03.1} the time--dependent averaged lifetimes (defined in \ocite{Sangalli2015}) at the $L$ and $L'$ high--symmetry points, $\oo{\gc}_{L/L'}\(t\)$ of bulk Silicon are
shown together with the equilibrium value. 

The first striking result is that, while the equilibrium lifetime is the same at the $L$ and $L'$ point (the two points are one the rotated of the other) the
NEQ lifetimes are not. This is due to the fact that the laser field breaks the rotational symmetry and induces a different population at the $L$ and $L'$
points. This, in turn, induce different NEQ lifetimes. A similar mechanism has been discussed in \ocite{PhysRevLett.114.247001} and \ocite{PhysRevX.8.041009}.  
It is also remarkable that the NEQ lifetimes at the $L'$ points becomes negative at early times. This is due to the fact that carriers are injected by the
experimental laser only in the $L$ point. A negative lifetime is sign of a super fast transfer of carriers from $L$ to $L'$. This fast process, that is entirely
due to a symmetry restoring dynamics, has been erroneously interpreted in \ocite{Ichibayashi2009} as an intrinsic process. Instead the simulations revealed that
it is instead entirely due to the experimental setup.

After the laser pulse the two NEQ lifetimes quickly tend to the same value restoring the rotation invariance. When the full relaxation is achieved the lifetimes
converge to a value that firmly remains below the equilibrium value. This demonstrates that the TR--ARPES induces effects whose interpretation must include the
photo--excitation process.

\section{Interacting electrons: Hamiltonian and potentials}
\lab{sec:H}
The theoretical description of the TR--ARPES experiments is bound to a correct definition of the different quantum objects involved in the dynamics.  These
objects can have an entire electronic character (plasmons, excitons...) or an independent bosonic character, like phonons. In case one is interested in
photoluminescence also photons must be treated quantum mechanically~\cite{Melo2016}. In this work we restrict the description to a classical treatment of the
electromagnetic field, replaced by the macroscopic external one.

We start, therefore, from a general non relativistic Hamiltonian describing a system of interacting electrons moving under the action of an external
electromagnetic field and of the internal electron--nucleus interaction. We will discuss in the following how this Hamiltonian can be simplified by using an
\ai\, approach and/or approximating some of its components.

This time dependent Hamiltonian can be split in the sum of a static, equilibrium term perturbed by the external time--dependent potential:
\eq{
\hat{H}\(t\)=\hat{H}^{eq}+\hat{V}\(t\).
\label{eq:full_h.0}
}
The external term, $\hat{V}$ is the one describing the interaction with the external electromagnetic field and will be discussed shortly. The equilibrium part of $\hat{H}\(t\)$ is composed
by free terms plus interaction potentials:
\seq{
\label{eq:full_h}
\eqg{
\hat{H}^0=\int\,d\rr \hat{\psi}^\dagger\(\rr\) \hat{\psi}\(\rr\)h\(\rr\)+ \hat{H}_p,\\
\hat{H}_\mathrm{int}=\hat{H}_\mathrm{e-e} + \hat{H}_\mathrm{e-p}.
}
}
\elab{eq:full_h}{a} represents the free electron and phonon terms. The first is written in terms of electronic field operators $\hat{\psi}$. These can be
expanded in a given orthonormal single--particle basis:
\eq{
\hat{\psi}\(\rr\)=\sum_{i}\vf_{i}\(\rr\)\hat{c}_{i}.
\label{eq:2nd_quant1}
}
Thanks to \e{eq:2nd_quant1} \elab{eq:full_h}{a} becomes
\eq{
 \hat{H}^0=\sum_i \gee_i \hat{c}^\dagger_i \hat{c}_i +  \sum_\mu\go_\mu \hat{b}_\mu^\dagger \hat{b}_\mu,
 \label{eq:h0}
}
with $\hat{c}_i^\dag$ and $\hat{b}_\mu^\dag$ electron and phonon creation operators. 

The electron--phonon interaction term definition easily follow in the $\hat{b}$, $\hat{c}$ representation:
\eq{
 \hat{H}_{e-p}=\sum_{ij\mu} g_{ij}^{\mu} \hat{c}^\dagger_i \hat{c}_j\(\hat{b}_\mu+\hat{b}^\dagger_\mu\),
 \label{eq:Hep}
}
with $g_{ij}^{\mu}$ the electron--phonon matrix elements~\cite{Marini2015,Giustino2017}. 

The electron--electron interaction is readily written in terms of electronic fields
\eq{
\hat{H}_{e-e} = \frac{1}{2}\int \rr \rr' \, \hat{\psi}^\dagger\(\rr\)\hat{\psi}^\dagger\(\rr'\)v\(\rr - \rr'\)\hat{\psi}\(\rr'\)\hat{\psi}\(\rr\),
\label{eq:h_e-e}
}
with $v\(\rr-\rr'\)$ the bare Coulomb potential. \e{eq:h_e-e} can be further expanded in the reference basis by using \e{eq:2nd_quant1}
\eq{
\hat{H}_{e-e} = \sum_{ijkl} v_{ijkl} \hat{c}_i^\dagger  \hat{c}_j^\dagger  \hat{c}_k \hat{c}_l.
\label{eq:h_e-e.1}
}
The last term to be explicitly written is the interaction with the external electromagnetic field.
As  the light--matter interaction is quadratic in the fermionic operators,  the most general time--dependent potential reads
\eq{
\hat{V}\(t\)=  \sum_{ij}V_{ij}(t)\hat{c}^{\dagger}_{i}\hat{c}_{j}.
\label{eq:Vt}
}
In \e{eq:Vt} $V_{ij}(t)$ is the matrix element of the single particle operator 
$\hat{V}(\blr,t)=\blA(\blr,t)\cdot\hat{\blj}(\blr,t)$ between the basis states $i$ and $j$.
Here $\blA(\blr,t)$ is the vector potential of the external electromagnetic field
and $\hat{\blj}(\blr,t)=[\hat{\blv}\,\delta(\hat{\blr}-\blr)+\delta(\hat{\blr}-\blr)\,\hat{\blv}]/2$
the current density operator, with $\hat{\blv}=\hat{\blp}+\blA(\hat{\blr},t)$.
In this work we assume to be in the dipole approximation, where $\blA(\blr,t)=\blA(t)$
and the coupling with the external field can be also expressed as
$\hat{V}(\blr,t)=\blE(t)\cdot\hat{\blr}$, with $\blE(t)$ the external field.

\subsection{The reference ab--initio system}
\lab{sec:AI_connection}
In the previous section  we have expanded the field operators in an arbitrary orthonormal basis.  In practical calculations, however, the basis must be
specified. In solids the index $i=n\blk\s$ comprises a  band-index $n$, a quasi-momentum $\blk$ and a spin orientation $\s$. 
This is the basis of Bloch states and it is particularly suited for  crystals. In finite systems like atoms and molecules a more convenient 
basis could be made of Slater-type orbitals, Gaussian-type orbitals, etc. 

The key property of this basis is that, by definition, it diagonalizes the single--particle and free Hamiltonian and defines the electronic single--particle energies
$\gee_i$. Thus the choice of the basis and reference single--particle Hamiltonian is crucial to embody some level of correlation already at the
independent--particle level. The question, now, is how to choose this basis. There are several options:
\begin{itemize}
 \item[$\cdot$] {\it Hartree--Fock} basis. This is accurate for atoms and small molecules where correlation represents a minor correction and the physics is
dominated by exchange interactions.
 \item[$\cdot$] {\it Density--Functional Theory} basis. HF is not a good approximation, in general,  for more 
complex, or correlated, materials, since HF completely neglects electronic correlation.
A better starting point for extended systems is the Hamiltonian coming from Density-Functional Theory (DFT) 
where the self-energy is replaced by the static and local Kohn-Sham (KS) potential $V^{KS}$. 
\end{itemize}

In the DFT case the single particle Hamiltonian appearing in \e{eq:full_h} becomes:
\eq{
\hat{h}\(\rr\)=\hat{T}_e+\hat{V}_n\(\rr\)+\hat{V}^{KS}\[\gr\(\rr\)\]\(\rr\).
\lab{eq:h_dft}
}
Despite the 
exact $V^{KS}$ potential is not known, simple and reliable approximations exist, in particular the 
so called Local--Density Approximation\,(LDA) and the Generalized--Gradient Approximation\,(GGA).

When the reference, single--particle Hamiltonian, is chosen to be \e{eq:h_dft} the whole MBPT machinery is labeled as
equilibrium {\it ai-MBPT}, \ai\, Many--Body Perturbation Theory.

If electronic states can be taken from DFT, in the {\it ai-MBPT}, phonons, needed to write \e{eq:Hep}, are taken from Density Functional Perturbation
Theory\,(DFPT)\cite{Gonze1997a,Stefano2001}. The procedure to embody in the \e{eq:Hep} the DFPT phonon energies and electron--phonon matrix elements is more
involved than the electronic case. It is, indeed, not possible to correct the bare  phonon Hamiltonian like in \e{eq:h_dft}. Nevertheless it is possible to
formally define families of diagrams that can be approximated by using DFPT, like described in \ocite{Marini2015}.

\subsection{A model Hamiltonian}
\lab{sec:model_H}
It is not always possible to use the full Hamiltonian as the \ai\, MBPT scheme can turn very demanding forcing to use specific approximations. In order to
investigate more advanced phenomena and get inspiration for extensions of the \ai\, scheme we have often used model Hamiltonian.

The model Hamiltonian considered in this work is
a one-dimensional tight-binding insulator of $\mathcal{L}$ sites, with one valence band $i=v$ and one conduction band $i=c$ separated by a direct
gap $\epsilon_{g}$~\cite{yang2012minimal}.  Since the formation of excitons is due to the attraction between a valence hole and a conduction electron we discard
the Coulomb interaction between electrons in the same band. For simplicity we also discard spin. By denoting with $\blk=k$ the one-dimensional momentum and by
$\epsilon^{\rm qp}_{i\blk}\equiv \epsilon_{ik}$ the band dispersion, the Hamiltonian of the insulator reads
\ml{
\hat{H}_{model}^{\rm eq}=\sum_{k}(\gee_{vk}\hat{v}^{\dagger}_{k}\hat{v}_{k}
+\gee_{ck}\hat{c}^{\dagger}_{k}\hat{c}_{k})-
U\sum_{k}\hat{c}^{\dagger}_{k}\hat{c}_{k}
+\frac{U}{\mathcal{L}}\sum_{k_{1}k_{2}q}\hat{v}^{\dagger}_{k_{1}+q}\hat{c}^{\dagger}_{k_{2}-q}
\hat{c}_{k_{2}}\hat{v}_{k_{1}},
\label{minmodham}
}
where $\hat{c}_{v\blk}\equiv \hat{v}_{k}$ and $\hat{c}_{c\blk}\equiv \hat{c}_{k}$ annihilates an electron of momentum $k$ in the valence ans band respectively,
and $U$ is the short-range interband repulsion.  The second term represents the interaction of a conduction electron with the positive background in the valence
band. For this model the ground state is obtained by filling all single-particle valence states with one electron. Therefore the number of protons is equal to
the number of valence electrons $\mathcal{L}$ in the ground state. 
Since we will use \e{minmodham} to study either the early stage of the long--time regime of the pump-induced dynamics \e{minmodham} ignores phonon--relaxation
and decoherence.

\section{Ab--Initio Many--Body Perturbation Theory}
\lab{sec:MBPT}
The Hamiltonian \e{eq:full_h.0} is composed of a static term, $\hat{H}^{eq}$ plus a time--dependent perturbation, $\hat{V}\(t\)$. The solution of the Many--Body
problem for $\hat{H}^{eq}$ already represents a challenging task and it requires some fundamental approximations that we will briefly review in the next
sections. At the same time the equilibrium problem already allows to introduce some key concepts that will remain valid out--of--equilibrium. The whole
theoretical scheme, indeed, even if turning more complicate because of the explicit time--dependence of the perturbation, preserves the conceptual structure
that emerges clearly at the equilibrium.

The first and fundamental concept that emerges from the solution of the equilibrium problem is the concept of trajectory. All excited state properties of  
the many--body system can be deduced from the dynamics of an electron traveling in the fully interacting material. This travel is described, mathematically,
from the single--particle Green's function, that in equilibrium system, can be written as a time--ordered
object~\cite{Mahan1990,Bruus2002,Abrikosov_Gorkov_Dzyaloshinski_2012,Leeuwen2013}:
\eq{
G_{ij}\(t,t'\)=\th\(t-t'\)G_{ij}^{>}\(t,t'\)+\th\(t'-t\)G_{ij}^{<}\(t,t'\).
\label{eq:Gto}
}
The physical interpretation of \e{eq:Gto} is given by the two components that governs the dynamics in the different time orderings:
the greater ($G^{>}$) and lesser ($G^{<}$) Green's function\,(GF) which
describe the propagation of an added electron and hole respectively. 
From the definition of $G$ it follows that
\seq{
\label{G><}
\eqg{
G^{>}_{ij}\(t,t'\)=-i\average{\hat{c}_{i}\(t\)\hat{c}^{\dagger}_{j}\(t'\)},\\
G^{<}_{ij}\(t,t'\)=i\average{ \hat{c}^{\dagger}_{j}\(t'\)\hat{c}_{i}\(t\)}.
}}
where operators depend on time according to the Heisenberg picture 
and the average $\average{\ldots}$ is performed over the initial ground 
state (or the initial many-body density matrix at finite temperature).

At the equilibrium the description of the electronic motion simplifies enormously as 
the time-ordered Green's function  $G$ depends only on the time difference:
\eq{
 G_{ij}\(t,t'\) \xLongrightarrow[]{equilibrium} G_{ij}\(t-t'\).
}
In practice this means that the GF can be Fourier transformed. This allows to expand it in a form, Lehmann representation, which clearly shows that connection
between trajectories and excited state properties. Indeed if we take \e{eq:Gto} and the its two components, \e{G><}, and consider the zero--temperature case to
simplify the math, it is straightforward to realize that
\eq{
G_{ij}\(\go\)=\sum_I\frac{f_i^I \(f_j^I\)^*}{\go-\tilde{\gee}_I+i 0^+\sign\(\tilde{\gee}_I-\mu\)},
\lab{eq:lehmann}
}
with $\mu$ the chemical potential and
\eq{
\tilde{\gee}_I=
\begin{cases}
E_0^N-E_I^{N-1}&\text{when}\, \tilde{\gee}_I<\mu,\\
E_I^{N+1}-E_0^{N}&\text{when}\, \tilde{\gee}_I\eqslantgtr\mu.\\
\end{cases}
\lab{eq:lehmann.1}
}
In \e{eq:lehmann} 
\eq{
 f_i^I =
\begin{cases}
\bra{N-1,I}\hat{c}_i\ket{N,0}&\text{when}\, \tilde{\gee}_I<\mu,\\
\bra{N,0}\hat{c}_i\ket{N+1,I}&\text{when}\, \tilde{\gee}_I\eqslantgtr \mu.\\
\end{cases}
\lab{eq:lehmann.2}
}
\erange{eq:lehmann}{eq:lehmann.2} clearly shows that the knowledge of the frequency dependent single--particle GF is directly connected to the exact one--body
excitation energies of the system. 

Now the problem is how to calculate $G$. In MBPT $G$ is the solution of a set of integro--differential equations that can be derived by using two different,
although equivalent, paths.
One is based on the
standard diagrammatic technique~\cite{ALEXANDERL.FETTER1971} which constructs approximations for the different terms of the theory
by using a geometrical and graphical approach.
An alternative approach is based instead on the equation of motion approach\cite{Strinati1988}.
This methods leads, both in the equilibrium and the out--of--equilibrium regimes, to a closed set of integro--differential equations that at the equilibrium
are known as Hedin's equations~\cite{Hedin19701}.

Both approaches lead to an equation of motion for $G$ written in terms of an electronic self--energy, $\gS$. This is the
Dyson equation 
\eq{
\uu{G}\(t-t'\)=\uu{g}\(t-t'\)+\int dt_{1}dt_{2}\,\uu{g}\(t-t_{1}\)\uu{\gS}\(t_{1}-t_{2}\)\uu{G}\(t_{2}-t'\),
\lab{eq:dys}
}
where $g$  is the bare GF and $\gS$ the time-ordered (correlation) self-energy  respectively.  In \e{eq:dys} we used underlined
quantities to indicate matrices in the $\(ij\)$ basis.

If now we take into account the equation of motion of $g$,  i.e., 
\eq{
\[i\frac{d}{dt}-\uu{h}\]\uu{g}\(t,t'\)=\delta\(t,t'\)\uu{1},
} 
we can  rewrite the Dyson equation as the following integro-differential  equation
\eq{
\[i\frac{d}{dt}-\uu{h}\]\uu{G}\(t-t'\)=\delta\(t-t'\)\uu{1}+ \int d\oo{t}\,\uu{\gS}\(t-\oo{t}\) \uu{G}\(\oo{t}-t'\),
\label{eomeq}
}
that will be particularly useful when we will move to the non--equilibrium regime. \e{eomeq} comes together with its adjoin~\cite{Strinati1988}.

\e{eq:dys} makes clear that all the complexity of the many--body system is embodied in the self--energy operator. In some sense this quantity plays the same
role of exchange--correlation potential of DFT that, if exact provides the exact ground state density, but that, in practice, is approximated. Compared to DFT
the great advantage of MBPT is that it offers a systematic way to approximate
$\uu{\gS}$, since its exact expression in terms of $G$ and the bare e--e and e--p interaction is, in principle, known~\cite{Strinati1988,Giustino2017}. In
practice it involves the definition of very involved and inter--connected objects (like the vertex function) that impose the adoption of approximations.
Thanks to the perturbative nature of the MBPT scheme it is always possible to ground these approximations on physical grounds. 

\subsection{Quasiparticles and lifetimes}
\begin{figure}[H]
\centering
\includegraphics[width=0.8\textwidth]{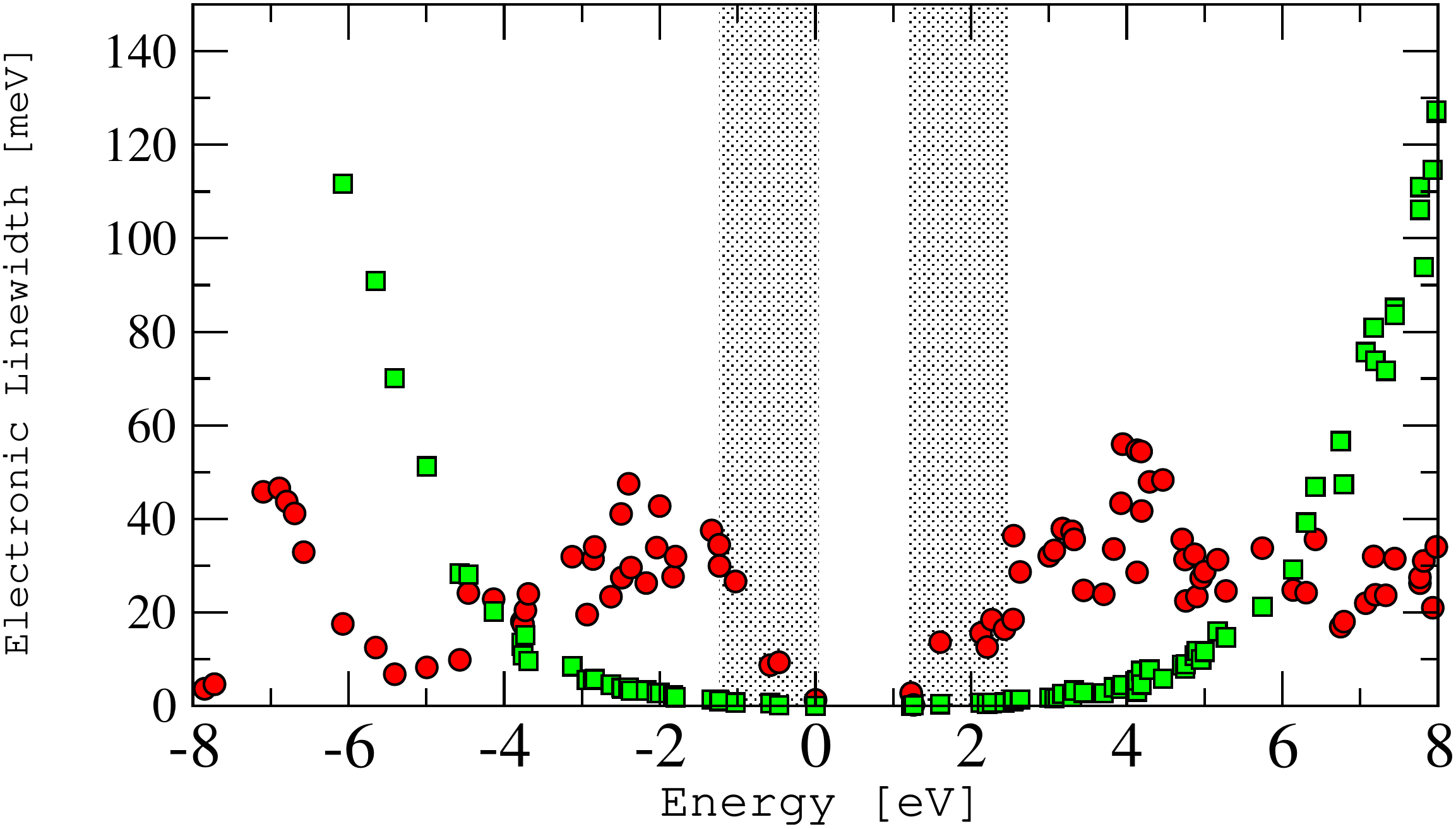}
\caption{Quasi-particle lifetimes in bulk Silicon. From Ref.~\cite{Marini2013}.
         Lifetimes are computes using \e{eq:Eqp} with the 
         electron--electron\,(boxes) and electron-phonon self-energy\,(circles).}
\label{fig:Si_eq_elph_lifetimes}
\end{figure}

The most elemental concept to use in order to show the power of the MBPT scheme is the quasiparticle\,(QP) concept. In DFT and in the KS scheme any electron is
accommodated on the levels corresponding to the eigenvalues of the KS Hamiltonian. These eigenvalues are, by construction, real. The MBPT gives a different and
alternative picture to it by distinguishing the different physical mechanism that are responsible of the dressing of the bare electron: classical,
exchange and, finally, correlation effects. All these effects are embodied in the self--energy and corresponds to specific approximations to it.

In order now to understand the effects induced by the frequency dependence of the correlated self--energy we can use
\e{eq:dys} and approximate $g$, $G$ and $\gS$ to be diagonal in the $i,j$ indexes. This allow a simple, analytic expression for the Dyson equation written in
frequency space
\eq{
G_i\(\go\)=g_i\(\go\)\[1+\gS_i\(\go\)G_i\(\go\)\].
\lab{eq:dysW}
}
From \e{eq:lehmann} we now know that 
\eq{
 g_i\(\go\)=\frac{1}{\go-\gee^0_i+i 0^+\sign\(\gee^0_I-\mu\)},
\lab{eq:g0}
}
with $\gee^0_i$ the bare, reference energies. Taking inspiration from \e{eq:g0} we can introduce the quasiparticle approximation by assuming that
\eq{
 G_i^{QP}\(\go\)\approx \frac{Z_i}{\go-\gee^{QP}}.
\lab{eq:Gqp}
}
If we plug \e{eq:Gqp} into \e{eq:dysW} we obtain
\eq{
\gee^{QP}_i = \gee^{0}_i + Z_i \gS_{ii}\(\gee^{QP}_i\),
\lab{eq:Eqp}
}
with $Z_i=(1-\partial_\go \gS|_{\go=\gee_i^{0}})^{-1}$ the renormalization factors. \e{eq:Eqp} contains a crucial difference compared to DFT and/or HF: the
QP energies $\gee^{QP}$ are complex. Physically we have that the real part of the QP energy corresponds to the renormalization of the single--particle energy
levels, while the imaginary part provides the QP width, that is the energy indetermination caused by quantistic effects. 

For reference in \fig{fig:Si_eq_elph_lifetimes}  we show the QP lifetimes (proportional to the inverse line-widths) in bulk Silicon from
\ocite{Marini2013}.
Both the electronic and the phonon--induced lifetimes are reported. Those are calculated in the $GW$ approximation that will be shortly reviewed in
\sec{sec:SEs}.

\subsubsection{Specific Self--Energy approximations}
\lab{sec:SEs}

\begin{figure}[H]
\centering
\includegraphics[width=0.5\textwidth]{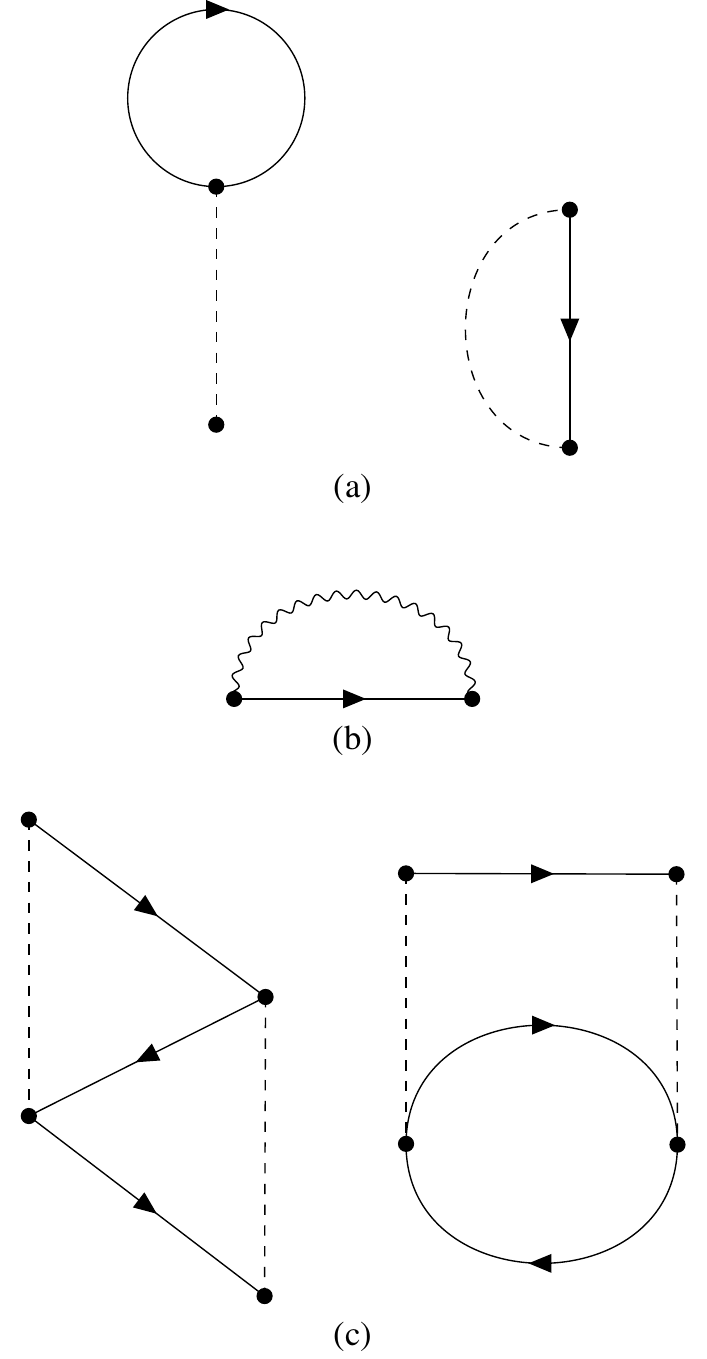}
\caption{Diagrammatic form of the self--energy approximations discussed in this work. We have the: (a) Hartree\,(left) and Fock\,(right), (b) $GW$, (c) second--Born
approximations. In the $GW$ approximation the wiggled propagator represent either the screened electronic interaction or the phonon propagator (see text). More
information can be found, for example, in \ocite{Leeuwen2013,Marini2013,Melo2016}.
}
\label{fig:SEs}
\end{figure}

{\em Hartree and Hartree Fock.}
At the lowest order of perturbation theory the two--body Coulomb interaction appearing in \e{eq:h_e-e.1} produces two terms, diagrammatically represented in
\figlab{fig:SEs}{a}:
\seq{
\lab{eq:hf}
\eqg{
\gS^{Hartree}_{ij}=-2i\sum_{\bar{m}\bar{n}}v_{i\bar{m}\bar{n} j}G_{\bar{n}\bar{m}}\(t,t^+\),\\
\gS^{Fock}_{ij}= i\sum_{\bar{m}\bar{n}}v_{i\bar{m} j\bar{n}}G_{\bar{n}\bar{m}}\(t,t^+\).
}
}
The first term is the Hartree or classical term with the factor 2 due to the spin summation. It can be easily visualized that it corresponds to solution of the Poisson equation by writing it in real space
\eq{
\gS^{Hartree}\(\rr,\rr'\)=-i\gd\(\rr-\rr'\)\int d\oo{\rr}v\(\rr-\oo{\rr}\)G\(\oo{\rr},\oo{\rr};t,t^+\),
\lab{eq:har}
}
with
\eq{
 G\(\rr,\rr';t,t'\)=\sum_{ij} \phi_i\(\rr\) \(\phi_j\(\rr'\)\)^*  G_{ij}\(t,t'\).
}
\e{eq:har} clearly shows that it represents the electrostatic potential generated by the electronic density, $\gr\(\oo{\rr}\)\equiv -i
G\(\oo{\rr},\oo{\rr};t,t^+\)$. A similar analysis can be done for the Fock term that reveals that it corresponds to exchange interaction caused by the
anti-symmetric properties of the electronic states.
As at equilibrium the GF depend only on the time difference the HF self--energy is static, $\uu{\gS}^{\rm HF}=\uu{\gS}^{Hartree}+\uu{\gS}^{Fock}$.

{\em The $GW$ approximation: the electronic case.}
All terms that go beyond the HF approximation are  commonly defined as
correlation terms.  The most striking difference with HF is that, even at equilibrium, the
correlation self--energy is frequency dependent reflecting the retardation effects caused by the interaction.

The general diagrammatic form of the $GW$ approximation is shown in \figlab{fig:SEs}{b} and it can be applied to either the electronic or the phonon mediated
interaction. 

We start, here, from the electronic case. 
As discussed both using the geometric, diagrammatic and the equation--of--motion approaches the most elemental excitation that is not included in the HF
approximation is the electron--hole pair creation. In practice it is possible to demonstrate that in the high density limit~\cite{ALEXANDERL.FETTER1971} all
correlation effects are exactly described by the charge oscillations described by the electronic response function, $\chi\(\rr,\rr';t-t'\)$. The physical
meaning of the response function can be easily visualized by noticing that it the same object entering the Kubo equation describing, within linear--response,
the reaction of a system to weak external perturbation. 

Let's, indeed, consider the potential defined in \e{eq:Vt} and let's also assume that $\hat{V}$ is so weak to be possible to be treated perturbatively. At the
first order this potential will induce a change in the density described by the Kubo equation
\eq{
\gd \gr\(\rr,t\)= \int dt'd\rr' \chi\(\rr,\rr';t-t'\) V\(\rr',t'\).
\lab{eq:kubo}
}
From \e{eq:kubo} we can also introduce an alternative definition of the response function, written as functional derivative,
\eq{
\chi\(\rr,\rr';t-t'\) = \frac{\gd \gr\(\rr,t\)}{\gd V\(\rr',t'\)}.
\lab{eq:drdv}
}
The next step we have to do in order to connect $\chi$ to correlation effects can be formally done~\cite{PhysRevB.38.2513} by using a variational approach.
Here, however, we follow a simpler approach entirely based on the \e{eq:kubo}. Indeed if we now we consider as perturbing potential $V\(\rr',t'\)$ the classical field
generated by a moving charge we readily see that \e{eq:kubo} describes the interaction of this charge with the electrons present in the material in terms of a
modified interaction
\ml{
W\(\rr,\rr';t-t'\) \equiv v\(\rr-\rr'\)\gd\(t-t'\)+\\  \int d\rr'\oo{\rr} v\(\rr-\oo{\rr}\)\chi\(\oo{\rr},\rr';t-t'\)v\(\rr-\rr'\)=\\
 \int d\rr' \gee^{-1}\(\oo{\rr},\rr';t-t'\)v\(\rr-\rr'\).
\lab{eq:W}
}
\e{eq:W} is the formal definition of the longitudinal screened interaction~\cite{Strinati1988} that can also be written as the product of the inverse dielectric
function $\gee^{-1}$ times the bare Coulomb interaction.
If we now consider the Fock self--energy we see that it represents exchange scatterings mediated by the bare e--e interaction. By using the simple dielectric
argument used to derive \e{eq:W} we deduce that the first effect of correlation will be transform the bare e--e interaction, in the Fock self--energy, in $W$.
Thus we have that 
\eq{
 \uu{\gS}^{Fock}_{ij} \xLongrightarrow[]{} \uu{\gS}^{GW}_{ij}= i\sum_{\bar{m}\bar{n}}\int d\go' W_{i\bar{m} j\bar{n}}\(\go-\go'\)G_{\bar{n}\bar{m}}\(\go'\).
 \lab{eq:GW}
}
It is worth to remind the \e{eq:GW} is exact in the homogeneous electron gas in the limit of high density. A condition, however, hardly satisfied in realistic
materials.

{\em The $GW$ approximation: the phononic case and the Fan approximation.}
When the $W$ propagator is replaced by the phonon propagator \e{eq:GW} defines another popular approximation very commonly used to describe
electron--phonon effects~\cite{Giustino2017}. In this case $W\rar D$:
\eq{
D_{\mu}\(t,t'\)=\th\(t-t'\)D_{\mu}^{>}\(t,t'\)+\th\(t'-t\)D_{\mu}^{<}\(t,t'\),
\label{eq:Dto}
}
with
\seq{
\label{D><}
\eqg{
D^{>}_{ij}\(t,t'\)=-i\average{\hat{q}_{\mu}(t)\hat{q}_{\mu}(t')},\\
D^{<}_{ij}\(t,t'\)=i\average{ \hat{q}_{\mu}(t')\hat{q}_{\mu}(t)},
}}
and $\hat{q}=\(\hat{b}_\mu+\hat{b}_\mu^\dag\)/\sqrt{2}$. When $W$ is replaced with $D$ in \e{eq:GW} we get the Fan self--energy:
\eq{
 \gS^{Fan}_{ij}\(t,t'\)=i\sum_{kl}\sum_{\mu} \(g^{\mu}_{ik}\)^* g^{\mu}_{jl} D_{\mu}\(t,t'\) G_{kl}\(t',t'\).
\label{eq:fan}
}
At the equilibrium $\gS$ and $D$ depend only on the time difference, as in \e{eq:GW}. 

The Fan self--energy, plus the Debye--Waller correction~\cite{Giustino2017} has been applied extensively to investigate the effect of the electron--phonon
interaction on the electronic and optical properties of a wealth of materials. One of the first applications was the simulation of the finite temperature
properties of extended systems~\cite{Marini2008}.

{\em The second--Born (2B) approximation.}
The driving mechanism that motivates the success of the $GW$ approximation is the description in terms of dielectric screening, $\x$. In systems composed of many
electrons this is the most elemental process that is also directly connected to the electrostatics of the material. When, however, the system is small and there
are not many electrons collective excitation, as those needed to build up screening, are not possible or less important. In these systems it is much more
important to take correctly into account exchange diagrams and avoid self--interaction errors.

For these kind of systems where screening is not dominant a well--known approximation is the  second--Born\,(2B). Diagrammatically this is represented by
\figlab{fig:SEs}{c} and it is composed of two terms. The first (right diagram) is in common with the $GW$ approximation and corresponds to the lowest order diagram. The second
(left diagram) is the exchange and, from the $GW$ perspective, represents the lowest order vertex of the self--energy.

The 2B self--energy can be written as
\eq{
\gS^{2B}_{ij}\(t,t'\)=
\sum_{nmpqsr}v_{irpn}w_{mqsj}G_{nm}\(t,t'\)G_{pq}\(t,t'\)G_{sr}(t',t),
\label{2bse}
}
where 
\eq{
w_{imnj}\equiv v_{imnj}-v_{imjn}.
}

\subsection{Neutral excitation: plasmons and excitons}
\lab{sec:eq_excitons}
The response function $\x\(\rr,\rr';\go\)$ is formally defined via the Kubo relation, \e{eq:kubo}, which can be rewritten in terms  of a density
variation,\e{eq:drdv}.  The Kubo equations, is however, a very general tool to describe the variation of any observable given the external perturbation.

In order to connect the response function  to the macroscopic, observable, physics we need however to carefully define the perturbing field. Indeed in realistic
materials we have three possibilities:
\begin{itemize}
 \item[(a)] The perturbing field is the external, bare field,
 \item[(b)] The perturbing field is the {\em total} field, i.e. the sum of the external plus the induced fields,
 \item[(b)] The perturbing field is the {\em total} and {\em macroscopic} field corresponding to the macroscopic component of the total field.
\end{itemize}
Each of the above definitions of the external perturbation defines, via the Kubo relation, a corresponding response function. 

\subsubsection{Potentials and response functions}
Let's start from case (a). In this case the definition of the perturbing field is simple and Kubo equation reduces to \e{eq:kubo} and \e{eq:drdv}. We have, then,
that upon application of the external perturbation the system will induce a variation of the density. If now we use simple electrostatic arguments we see
that this variation will correspond to a variation of Hartree potential, solution of Poisson equation:
\eq{
 \gd\gS^{Hartree}\(\rr,t\)= \int\d\rr' v\(\rr-\rr'\)\gd\gr\(\rr',t\),
 \lab{eq:Vind}
}
\e{eq:Vind} defines the {\em induced potential}. This, physically, corresponds to the potential induced by the external perturbation. 

At this point we have two potentials: an external $V\(\rr,t\)$, time--dependent and spatially homogeneous, and a microscopic, $\gd\gS^{Hartree}\(\rr,t\)$,
time--dependent but with microscopic components.

In an optical absorption experiment the scattered light measured corresponds, for obvious reasons, to the total and {\em macroscopic} potential. The total
potential is the sum of $V$ and $\gd\gS^{Hartree}$. In the macroscopic component, $V^{tot}_M$ is spatially averaged as the external potential is, by definition,
macroscopic:
\seq{
 \lab{eq:Vmac}
\eqg{
 V^{tot}\(\rr,t\)=V\(\rr,t\)+\gd\gS^{Hartree}\(\rr,t\),\\
 V^{tot}_{M}\(\rr,t\)=V\(\rr,t\)+\frac{1}{\Omega}\int d\rr \gd\gS^{Hartree}\(\rr,t\),
}}
with $\Omega$ the crystal volume.

We have now the formal definition of the three possible perturbing fields. For each of them \e{eq:kubo} defines a response function. 
In addition to $\chi$, therefore, we have
\seq{
\lab{eq:chis}
\eqg{
 \chi^{tot}\(\rr,\rr';t-t'\) = \frac{\gd \gr\(\rr,t\)}{\gd V^{tot}\(\rr',t'\)},\\
 \chi_{M}\(\rr,\rr';t-t'\) = \frac{\gd \gr\(\rr,t\)}{\gd V^{tot}_M\(\rr',t'\)}.
} 
}
The natural question is: which one is the correct one to be used to interpret a given experiment? 
In the case of the optical absorption we know that the macroscopic absorption is given by
\eq{
 \gee_M\(\go\)=\frac{1}{\average{\average{\gee^{-1}\(\rr\,\rr';\go\)}}}=1-\average{\average{\int d\oo{\rr} v\(\rr-\oo{\rr}\)\chi_{M}\(\oo{\rr},\rr';\go\)}}.
 \lab{eq:eps2}
}
Similarly the electron--energy loss spectrum is given by
\eq{
 \gee^{-1}\(\go\)=\average{\average{\gee^{-1}\(\rr\,\rr';\go\)}}=1+\average{\average{\int d\oo{\rr} v\(\rr-\oo{\rr}\)\chi\(\oo{\rr},\rr';\go\)}},
 \lab{eq:eels}
}
where we have introduced the double brackets notation to represent a macroscopic average:
\eq{
 \average{\average{f\(\rr,\rr'\)}}\equiv \int d\rr d\rr' f\(\rr,\rr'\).
}
From  \e{eq:eps2} and \e{eq:eels} it follows that we need different response function depending on the experimental setup. 

A last remark is that, by following the same strategy adopted in the single--particle GF case, also $\x$ can be represented in a Lehmann form. The form shows that
the poles of $\x$ are number conserving excitation $E_I(N)-E_0(N)$. 

\subsubsection{The Bethe--Salpeter equation(s)}
\lab{sec:bse}
It is well known\cite{Onida2002} that the response function satisfies an equation named Bethe--Salpeter Equation\,(BSE). From \e{eq:chis} is evident that we have more than one
BSE depending on which response function (observable) we are interested in. 

Let's take as example $\x$. $\x$ can be derived, in the framework of MBPT, from the four point electron--hole propagator $\x_{ijkl}\(\w\)$. This is defined by expanding 
 \e{eq:drdv} in the reference basis
\eq{
 \x_{ijkl}\(t-t'\)\equiv\frac{\gd \gr_{ij}\(t,t^+\)}{\gd V_{lk}\(t'\)}.
 \lab{eq:dGdV}
}
While $\x$ is a two points function in real space $\x$ is a four point function in the single--particle basis~\cite{Strinati1988}.

The equation of motion for $\chi$ follows exactly the same path we outlined in the case of the single--particle GF,
\seclab{sec:MBPT}. We can use either the diagrammatic path or the equation of motion approach.
From \e{eq:dGdV} we observe that $-i G_{ij}\(t,t^+\)=\gr_{ij}\(t\)$, with $\uu{\gr}$ the density matrix. Now, the equation of  motion for $\gr$ follows from
the equal time contraction of the \e{eomeq}. Let's consider, for the moment, just the HF contribution to the self--energy operator. This term is time
independent and simplifies the math
\eq{
 \frac{d}{dt} \uu{\gr}\(t\)=-i\[\uu{h}+\uu{\gS}^{Fock}\(t\)+\uu{V}^{tot}\(t\),\uu{\gr}\(t\)\].
 \lab{eq:drhodt}
}
In \e{eq:drhodt} $\gS^{HF}$ is the time--dependent HF self--energy, which is functional of the density matrix itself, see \e{eq:hf}.
From \e{eq:drhodt} the equation of motion for $\chi$ follows easily.
\eq{
  \frac{d}{dt}\uuu{\x}\(t-t'\)=-i\[ \uu{h},\uuu{\x}\(t-t'\)\]
  -i\[
  \uuu{K}^{Fock}\circ\uuu{\x}\(t,t'\)+
  \frac{\gd\uu{V}^{tot}\(t\)}{\gd \uu{V}\(t'\)}
  ,\uu{\gr}_0\],
 \lab{eq:BSE}
}
with
\eq{
\gd \uu{\gS}^{Fock} = \left.\uuu{K}^{Fock} \circ \gd\uu{\gr}\(t\)\right|_{\gr=\gr_0}.
}
\e{eq:BSE} is the BSE for the generic response function $\x$.
In \e{eq:BSE} we have also introduced a short notation for matrix--matrix and matrix--vector multiplications. If $M$ and $N$ are
two matrices and $V$ is a vector we define
\seq{
\lab{eq:BSE_not}
\eqg{
(M\circ V)_{pq}\equiv M_{\substack{ pq\\mn}}V_{nm}, \\
(M \circ N)_{\substack{ mn\\ rs}}\equiv M_{\substack{ mn\\pq}}  N_{\substack{ qp\\rs}}, \\
[N , V]_{\substack{ mn\\ pq}}=-[V , N]_{\substack{ mn\\ pq}}\equiv N_{\substack{ mi\\pq}} V_{in} -  V_{mi}  N_{\substack{ in\\pq}}.
}}

In order to make explicit the last term of \e{eq:BSE} we need to specify which polarization function we are interested in. Indeed, 
the final form of the BSE depends 
depends on the definition of the perturbing field. Indeed we have that
\eq{
 \frac{\gd\uu{V}^{tot}\(t\)}{\gd \uu{V}\(t'\)}=
 \begin{cases}
   \left.\frac{\gd\uu{\gS}^{Hartree}\(t\)}{\gd \uu{V}\(t'\)}\right|_{\gr=\gr_0}+\uuu{1}\gd\(t-t'\) &  \\
   \uuu{1}\gd\(t-t'\) & \x=\x^{tot}\\
   \left.\frac{\gd\uu{V}^{tot}_M\(t\)}{\gd \uu{V}\(t'\)}\right|_{\gr=\gr_0} & \x=\x^{tot}_M
 \end{cases}
 .
}

\section{{\it Ab--Initio} Non--Equilibrium Green's Function Theory}
\lab{sec:NEGF}
So far we have dealt with systems in equilibrium where the techniques of MBPT applies. This approach, however, is based on some key assumptions. The system is
supposed to: (i) be time invariant, i.e. all observable depend only on time differences; (ii) be in the ground state and/or in a thermally distributed mixture
of excited states.

These two assumptions clearly fail in the case where an external time--dependent perturbation is applied. When this perturbation is weak it is still possible to
describe the system in terms of the equilibrium Hamiltonian and treat the perturbation within Linear--Response regime. 

In typical \ppe\, experiments, however, the external perturbation can be arbitrary strong and the two conditions on which equilibrium MBPT is based chase to
hold. Nevertheless, in this regime, it is possible to extend the  Green's function approach as well as the approximations generated from MBPT. This extension
is the Non--Equilibrium Green Function approach\,(NEGF). The NEGF scheme has been devised in the sixties and it is reviewed, for example, in
Ref.\cite{Leeuwen2013}.

\subsection{NEGF and connection to observable quantities}
Let's indeed consider now the action of the external perturbation of \e{eq:full_h.0}. The external driving field $V\(t\)$ breaks the invariance under
time-translations. Consequently any observable acquires and explicit time dependence. Let's take as an example  the HF self-energy,
\eq{
\gS^{\rm HF}_{ij}\(t\)=\sum_{\bar{m}\bar{n}}\(v_{i\bar{m}\bar{n} j}-v_{i\bar{m} j\bar{n}}\)\r_{\bar{n}\bar{m}}\(t\),
}
whose time dependence is trough $\r_{nm}\(t\)$, the one-particle density matrix.
We can thus define the time dependent NEQ Hamiltonian:
\eq{
\uu{h}^{\rm neq}\(t\)=\uu{h}+\uu{\gS}^{\rm HF}\(t\)+\uu{V}\(t\)=\uu{h}^{\rm HF}\(t\)+\uu{V}\(t\).
\lab{eq:H_MB_def_neq}
}

The $G^{\lessgtr}$ functions in Eq.~(\ref{G><}) contains a wealth of information  about the non-equilibrium system. From the {\em equal-time} $G^{<}(t,t)$ we 
have access to the one-particle density matrix according to
\eq{
 \uu{\r}\(t\)=-i\uu{G}^{<}\(t,t\),
}
and therefore any time-dependent average of any one-body operator
$\hat{O}=\sum_{ij}O_{ij}\hat{c}^{\dagger}_{i}\hat{c}_{j}$ can be evaluated as:
\eq{
O\(t\) =  \sum_{ij}O_{ij}\average{\hat{c}^{\dagger}_{i}(t')\hat{c}_{j}\(t\)} =-i\sum_{ij}O_{ij}G_{ji}^{<}(t,t).
}
Thus, $G^{<}(t,t)$  can be used to calculate time-dependent local densities, currents, spin-densities, spin-currents, orbital magnetic moments, dipole moments,
etc. Interestingly, for weak perturbations $V$ the Fourier transform of $O\(t\)$ is peaked at the absorption energies of the equilibrium system. Hence,
real-time simulations provide an alternative route to obtain different kind of response functions.  Knowledge of the {\em off-diagonal} (in time)
$G^{<}_{ij}\(t,t'\)$ allows for the calculation of the time-dependent total energy~\cite{Leeuwen2013}, i.e., the average of $\hat{h}^{\rm neq}\(t\)$ defined in
Eq.~(\ref{eq:H_MB_def_neq}):
\ml{
E\(t\)\equiv\average{\hat{h}^{\rm neq}\(t\)}=
-\frac{i}{2}\sum_{ij}\left(h^{\rm HF}_{ij}\(t\)+V_{ij}\(t\)\right)G^{<}_{ji}\(t,t\)=\\
-\frac{i}{4}\sum_{i}\left[\left(i\frac{d}{dt}-i\frac{d}{dt'}\right)G^{<}_{ii}\(t,t'\)\right]_{t=t'}.
}
Finally, the Fourier transform of $G^{\lessgtr}\(t,t'\)$ with 
respect to the relative time $t-t'$ yields the time-resolved spectral 
function
\eq{
A^{\lessgtr}\(T=\frac{t+t'}{2},\w\)\equiv -i\int d\(t-t'\) e^{i\w\(t-t'\)}\mathrm{Tr}[\uu{G}^{\lessgtr}\(t,t'\)].
}
The quantity $A^{<}$ ($A^{>}$) is peaked at the removal (addition) single-particle energies of the system at time $T$ and hence it can be used to address
transient photo-emission spectra (inverse photo-emission spectra) as we shall see later.

Given the importance of the lesser/greater Green's functions in the characterization of a physical system we now discuss briefly how they can be calculated in
practice.

\subsection{From the full Kadanoff-Baym equations to the single--time evolution method}
\lab{sec:GKBA}
In non-equilibrium the equation of motion \e{eomeq} is replaced by  a coupled system of integro-differential equations for the lesser/greater 
Green's functions, also known as the {\em Kadanoff-Baym equations}\,(KBE)~\cite{Kadanoff1962,Haug1996,SDvL.2009,vLDSAvB.2006,Leeuwen2013,d.1984}
\seq{
\lab{kbe}
\eqg{
\[
i\frac{d}{dt}\uu{1}-\uu{h}^{\rm HF}\(t\)\right]\uu{G}^{<}\(t,t'\)=\uu{I}^{<}\(t,t'\),\\ 
\uu{G}^{>}\(t,t'\)\left[ -i\frac{\overleftarrow{d}}{dt'}\uu{1}-\uu{h}^{\rm HF}(t')\]=\uu{I}^{>}\(t,t'\),
}
}
with collision integrals
\seq{
\lab{collint}
\eqg{
\uu{I}^{<}\(t,t'\)=
\int_{-\infty}^{\infty} dt_{1}\left[\uu{\gS}^{<}\(t,t_{1}\)\uu{G}^{\rm A}\(t_{1},t'\)+ \uu{\gS}^{\rm R}\(t,t_{1}\)\uu{G}^{<}\(t_{1},t'\)\right],\\
\uu{I}^{>}\(t,t'\)= \int_{-\infty}^{\infty} dt_{1}\left[\uu{G}^{>}\(t,t_{1}\)\uu{\gS}^{\rm A}\(t_{1},t'\)+ \uu{G}^{\rm R}\(t,t_{1}\)\uu{\gS}^{>}\(t_{1},t'\)\right].
}}
In \e{collint} it appears the retarted/advanced 
component of the Green's function and self-energy. They are defined 
according to
\eq{
X^{\rm R}\(t,t'\)=[X^{\rm A}(t',t)]^{\dagger}=
\th(t-t')\left[X^{>}\(t,t'\)-X^{<}\(t,t'\)\right],
\label{xr}
}
where $X$ can be any other two-time correlator.  The collision  integrals $I^{\lessgtr}$ encode all correlation effects via the many-body self-energy $\gS_{c}$.
Since $\gS^{\lessgtr}=\gS^{\lessgtr}[G^{>},G^{<}]$ are  functionals of $G^{>}$ and $G^{<}$, \e{kbe} constitute a closed nonlinear system of integro-differential
equations. We refer to this approach of tackling non-equilibrium problems as the {\em Non-equilibrium Green's function} (NEGF) method.

From Eqs.~(\ref{kbe},\ref{collint}) it is evident that the computational cost of solving the NEGF equations scales cubically with the propagation time.  For
this reason the full numerical solution of the KBE has so far being limited to light atoms and diatomic molecules~\cite{DvL.2006,HBBB.2010,BBB.2010,BBB2.2010},
the homogeneous electron gas~\cite{KB.2000,GMSBK.2003,SBK.2003}, the two-band model~\cite{HS.1992,BKSBKK.1996,KB.1997,BKB.1997,KBBK.1999} and several model
Hamiltonians for finite systems~\cite{BBvLSD.2009,vFVA.2010,SMvL.2012,BHB.2012,SBP.2016}, lattice systems~\cite{FTZ.2006,EKW.2010,EW.2010} and open
systems~\cite{MSSvL.2008,MSSvL.2009,vFVA.2009,MSSvL.2010,KLAF.2011,UKSSKvLG.2011,KUSSKvLG.2012,PS.2013,KV.2014}.

In order to allow for first-principles NEGF simulations of complex molecules or crystals, one has to circumvent the problem of the cubic scaling.  A way to
drastically reduce the computational time consists in implementing the so-called Generalized Kadanoff-Baym Ansatz (GKBA)~\cite{LSV.1986}.  The basic idea is to
collapse the KBE into a single equation for the one-particle density matrix $\r\(t\)=-iG^{<}(t,t)$ from which one can calculate the time-dependent averages of all
one-body observable. Thus, with the GKBA we loose the information brought by the off-diagonal in time Green's functions but we gain in computational
efficiency.  

The exact equation for $\r\(t\)$, that was derived in \sec{sec:bse} within the HF approximation, can be also derived within NEGF from the difference between \elab{kbe}{a} and \elab{kbe}{b}, and reads 
\eq{
\frac{d}{dt}\uu{\r}\(t\)+i\left[\uu{h}^{\rm HF}\(t\),\uu{\r}\(t\)\right]=-\uu{I}\(t\) ,
\label{eomrho}
}
with 
\eq{
\uu{I}\(t\)\equiv \uu{I}^{<}(t,t)+{\rm H.c.}
}
This is not a closed equation for $\r$ since the collision integral 
\ml{
\uu{I}\(t\)=-\int_{-\infty}^{t}dt'\[\uu{\gS}^{<}\(t,t'\)\uu{G}^{>}\(t',t\)+ \uu{\gS}^{>}\(t,t'\)\uu{G}^{<}\(t',t\)\nl
+\uu{G}^{<}\(t,t'\)\uu{\gS}^{>}\(t',t\)+ \uu{G}^{>}\(t,t'\)\uu{\gS}^{<}\(t',t\)\].
\label{kcoll}
}
depends on the off-diagonal in time $G^{\lessgtr}$ both explicitly and 
implicitly through $\gS$. 

According the the GKBA one can close Eq.~(\ref{eomrho})  by approximating
the lesser and greater Green' function as  
\seq{
\label{gkba}
\eqg{
\uu{G}^{<}\(t,t'\)=-\uu{G}^{\rm R}\(t,t'\)\uu{\r}\(t'\)+\r\(t\)\uu{G}^{\rm A}\(t,t'\),\\
\uu{G}^{>}\(t,t'\)=\uu{G}^{\rm R}\(t,t'\)\uu{\bar{\r}}\(t'\)-\uu{\bar{\r}}\(t\)\uu{G}^{\rm A}\(t,t'\),
}}
where $\bar{\r}\(t\)=1-\r\(t\)$, and the quasi-particle propagator $G^{\rm R}$ (and hence $G^{\rm A}=\[G^{\rm R}\]^{\dagger}$) as
\eq{
\uu{G}^{\rm R}\(t,t'\)=-i\th\(t-t'\)Te^{-i\int_{t'}^{t}d\bar{t}\,\uu{h}^{\rm QP}(\bar{t})}.
\label{gret}
}
For (small) finite systems the choice $h^{\rm QP}=h^{\rm HF}\(t\)$ is a good choice~\cite{PUvLS.2015}. For extended systems, however, the lack of damping in
$h^{\rm eq}$ prevents the system to relax.  In these cases the propagator is typically corrected by adding  non--hermitian terms given by the quasi--particle
life--times, see \sec{sec:SEs}. In this case the self--energy is not hermitian and the quasi--particle energy includes an imaginary part~\cite{Marini2008,ALA.2005,marini.2003,h.1992,bsh.1999,m.2013,LPUvLS.2014}.
\eq{
h_{ij}^{\rm QP}=\gd_{ij}\[\Re\(\gee^{QP}_i\)+i\Im\(\gee^{QP}_i\)\].
}

To summarize, Eq.~(\ref{eomrho}) together with  Eq.~(\ref{gkba}) and  Eq.~(\ref{gret}) constitute a closed equation of motion for $\rho$ called {\it GKBA
method}. The computational cost of this method scales quadratic with the propagation time (if the calculation of the self-energy does not scale worse than
it), thus offering a huge gain with respect to the solution of the KBE.

Among the properties of the GKBA we mention the fulfillment of the relation $G^{\rm R}-G^{\rm A}=G^{>}-G^{<}$ for any choice of $G^{\rm R}$, and the fact that
Eqs.~(\ref{gkba}) become an identity in the limit $t\to t'$ since $G^{\rm R}(t^{+},t)=-i$. Another important feature  is that the GKBA preserves the continuity
equation. 

There is, however, an even more important property which clarifies the physics contained in the GKBA.  In the HF approximation the collision integral
vanishes and \e{gkba} are the solution of \e{eomrho} provided that  $G^{\rm R}$ is the HF propagator.  Therefore, the more the quasi-particle picture is
valid the more the GKBA is accurate.  This is typically the case when the time between two consecutive collisions is smaller than the quasi-particle life-time.

The GKBA approach has been successfully applied in many contexts, including both finite and extended systems.  Recent applications include the non-equilibrium
dynamics~\cite{Hermanns2014, schlunzen2016} and many-body localization~\cite{lev2014} of Hubbard clusters, time-dependent quantum
transport~\cite{LPUvLS.2014,hopjan2018molecular,Riku2021,cosco2020spectral}, real-time description of the Auger
decay~\cite{covito2018benchmarking}, transient absorption of atoms~\cite{PUvLS.2015,perfetto2016time}, ultrafast charge migration in
bio-molecules~\cite{perfetto_ultrafast_2018,perfetto_first-principles_2019,perfetto_ultrafast_2020,mansson2021correlation} and donor/acceptor
dyads~\cite{bostrom_charge_2018}, time-resolved spectroscopy of excitonic insulators out of
equilibrium~\cite{murakami_ultrafast_2020,doi:10.1002/pssb.201800469,tuovinen_comparing_2020}, equilibrium absorption of sodium
clusters~\cite{pal_optical_2011}, transient absorption~\cite{PSMS.2015,SDCMCM.2016,Pogna2016} and carrier dynamics~\cite{Sangalli2015,Molina-Sanchez2017} of
semiconductors.

Below we discuss a possible next level of approximation to further reduce the computational effort of the GKBA method.

\subsection{The Markovian approximation}
\lab{sec:MA}
Through the GKBA, the self-energy functional $\gS^{\lessgtr}$ turn into functional of the retarded Green's function and one-particle density matrix, i.e.,
$\gS^{\lessgtr}\[G^{>},G^{<}\]\to \gS^{\lessgtr}\[G^{\rm R},\r\]$.  Using the GKBA also in \e{kcoll} the collision integral becomes:
\eq{
\uu{I}\(t\)=-\int_{-\infty}^{t}dt'  \mathcal{F}\Big[\uu{\gS}^{\lessgtr}\(t',t\)\[\uu{G}^{\rm R},\uu{\r}\],\uu{G}^{\rm R}\(t,t'\),\uu{\r}\(t'\)\Big] 
\label{collintGKBA}
}
i.e. $I\(t\)$ is an implicit functional, via $\gS^{\lessgtr}$, and also an explicit functional
of the whole history of $\r\(t'\)$ for $t'<t$. 
The {\em Markovian approximation}\,(MA) consists in neglecting memory effects, thus evaluating \e{collintGKBA} at the instantaneous density matrix
$\r\(t\)$:
\eq{
\uu{I}^{\rm MA}\(t\)=-\int_{-\infty}^{t}dt'  \mathcal{F}\Big[\uu{\gS}^{\lessgtr}\(t',t\)\[\uu{G}^{\rm R},\uu{\r}\(t\)\],\uu{G}^{\rm R}\(t,t'\),\uu{\r}\(t\)\Big] .
\label{collintMA}
}
where now $I^{\rm MA}$ is a functional of the instantaneous density.  The MA is justified provided that the electron dynamics is slow on the time-scale over
which the self-energy decays.

If we further assume that the effects of a change in the density matrix on the quasi-particle Hamiltonian $h^{\rm qp}\(t\)$ are only minor then we can use the
equilibrium value $h^{\rm qp}_{0}$ (which is independent of time) to evaluate the propagator:
\eq{
\uu{G}^{\rm R}\(t,t'\)=-i\th\(t-t'\)e^{-i\uu{h}^{\rm qp}_{0}\(t-t'\)}.
\label{greteq}
}
Assuming  $h^{\rm qp}_{ij}=\delta_{ij}\epsilon^{\rm QP}_{i}$ (see \sec{sec:SEs}), the  propagator gets the simpler expression
\eq{
G_{ij}^{\rm R}\(t,t'\)=-i\delta_{ij}\th(t-t')e^{-i\epsilon^{\rm QP}_{i}(t-t')}.
\label{greteq2}
}
With these approximations the time integral in $I\(t\)$ can be done analytically with the integrand depending only on the time difference $t-t'$:
\eq{
\uu{I}^{\rm MA}\(t\)\approx -\int_{-\infty}^{t}dt'\mathcal{\uu{K}}[\uu{\r}\(t\)]\(t-t'\).
\label{collintMA+QP}
}
Where the $\mathcal{K}$ functional is constructed explicitly using Eq.~(\ref{greteq}) in the definition of the $\mathcal{F}$ functional.  The dependence on
$t-t'$ stems from the time-translational invariance of $G^{\rm R}$ and it is therefore given by simple exponential functions.

\subsection{Equilibrium self--energies taken out--of--equilibrium}
\lab{sec:MA_self_energies}
All self--energies defined in \sec{sec:SEs} can be taken out--of--equilibrium by using NEGF techniques~\cite{Leeuwen2013}. The procedure is to take all time
arguments on the Keldysh contour and apply Langreth rules (see Chapter 5 of \ocite{Leeuwen2013}). 

The procedure is tedious but straightforward. In some cases, like the electronic $GW$ approximation, the self-energy in the out--of--equilibrium regime
acquires a different form compared to the equilibrium case. 

In this section we illustrate the MA using the 2B  self-energy, \e{2bse}. The electronic and phononic $GW$ self--energies written out--of--equilibrium by using the GKBA and MA can
be found, for example, in \ocite{Marini2013,Melo2016}.
In this case the self--energy
is an instantaneous functional of $\uu{G}^{\lessgtr}\(t,t'\)$
\eq{
\gS_{ij}^{\lessgtr}\(t,t'\)=\!\!\!
\sum_{nmpqsr}v_{irpn}w_{mqsj}G_{nm}^{\lessgtr}\(t,t'\)G_{pq}^{\lessgtr}\(t,t'\)
G^{\gtrless}_{sr}(t',t),
\label{2bse_ma}
}
where 
\eq{
w_{imnj}\equiv v_{imnj}-v_{imjn}.
}
Using the GKBA expression for the lesser and greater Green's function and by replacing the pair $\r(t')$ and $\bar{\r}(t')$ with the pair $\r\(t\)$ and
$\bar{\r}\(t\)$  the collision integral can be rewritten as
\ml{
I^{\rm MA}_{ik}\(t\)=\sum_{nmpqsr}\sum_{j}v_{irpn}w_{mqsj}\int_{-\infty}^{t} dt'\\
\[
\left(G^{\rm R}\(t,t'\)\r\(t\)\right)_{nm}
\left(G^{\rm R}\(t,t'\)\r\(t\)\right)_{pq}
\left(\bar{\r}\(t\)G^{\rm A}(t',t)\right)_{sr}
\left(\bar{\r}\(t\)G^{\rm A}(t',t)\right)_{jk}
\nl
-\left(G^{\rm R}\(t,t'\)\bar{\r}\(t\)\right)_{nm}
\left(G^{\rm R}\(t,t'\)\bar{\r}\(t\)\right)_{pq}
\left(\r\(t\)G^{\rm A}(t',t)\right)_{sr}
\left(\r\(t\)G^{\rm A}(t',t)\right)_{jk}\] \\
= \sum_{nmpqsr}\sum_{j}v_{irpn}w_{mqsj} 
\[\r_{nm}\(t\)\r_{pq}\(t\)\bar{\r}_{sr}\(t\)\bar{\r}_{jk}\(t\)-
\bar{\r}_{nm}\(t\)\bar{\r}_{pq}\(t\)\r_{sr}\(t\)\r_{jk}\(t\)\] \\
\times \int_{-\infty}^{t}  dt'  
e^{-i(\epsilon^{\rm qp}_{n}+\epsilon^{\rm qp}_{p}-\epsilon^{\rm 
qp}_{r}-\epsilon^{\rm qp}_{k})(t-t')}.
\label{collgkba_2b}
}
From \e{collgkba_2b} it is evident that  the collision integral is a quartic polynomial in $\r\(t\)$. As anticipated, the integral over $t'$ can be done
analytically due to the simplicity of the propagator in \e{greteq2}. Thus the equation of motion for $\r$ reduces to a  nonlinear differential equation
with temporal scaling reduced from quadratic to linear, by maintaining the {\it same scaling with the system size}.  It is worth noting that it has been
recently shown that linear scaling in time can be achieved within GKBA  without introducing further approximations, but at the cost of worsening the scaling
with the system size~\cite{schlunzenachieving2020,joostg1g22020,pavlyukh2021photoinduced}.


\section{Time--Resolved ARPES}
In time-resolved and angle-resolved photo-emission (tr-ARPES) experiments on semiconductors or insulators a pump pulse excites electrons from the valence band to
the conduction band. The photocurrent is generated by a {\em probe} pulse which impinges the system and causes the emission of electrons from the instantaneously
filled states. Below we present an expression for the momentum resolved photocurrent in terms of the lesser Green's function of the solid.  Here we assume that
the system is translationally invariant with quasi momentum $\blk$ and therefore it is useful to re-expand explicitly the index $i$ used so far as $i \to i\blk
\sigma$.  In particular  $\hat{c}_{i \blk \s} ^{(\dag)}$ annihilates (creates) an electron of momentum $\blk$ with spin $\s$ in band $i$ and energy
$\epsilon^{\rm qp}_{i\blk}$, and the corresponding lesser and greater Green's function in Eq.~\ref{G><} become
\seq{
\eqg{
G^{<}_{ij\blk}\(t,t'\)=i\langle  \hat{c}^{\dag}_{j \blk \s  }\(t'\)  \hat{c}_{i \blk \s}\(t\)   \rangle, \\
G^{>}_{ij\blk}\(t,t'\)=-i\langle  \hat{c}_{i \blk \s  }\(t\)  \hat{c}^{\dag}_{j \blk \s}\(t'\)   \rangle .
}
}
Similarly we have that $\mu \to \mu\qq$, with $\qq$ transferred momentum and
\seq{
\lab{eq:k_form}
\eqg{
 g^{\mu}_{ij} \Rar g^{\qq\mu}_{\kk ij},\\
 v_{ijml}\Rar v^\qq_{\substack{ ik\kk\\mk\pp}}.
}}
In all above definitions we assumed for simplicity that the material is invariant under spin rotations. The generalization to non spin-compensated  systems is
straightforward. In the case where the pump and probe fields are overlapping in time explicitly gauge--invariant expressions for the photo--current have been 
proposed\cite{Freericks_2015,PhysRevB.44.3655}. Nevertheless, as explained in \sec{sec:H}, in this work we assume to be in the dipole approximation.  Therefore 
the full Hamiltonian in the
presence of the pump field is itself gauge invariant as only the gauge-invariant electric field enters.
Consequently, in the dipole approximation adopted here no extra terms has to be added to restore the gauge invariance.

The tr-ARPES signal is proportional to the number of electrons $N_{\blk}(\w)$  with energy $\w$ and parallel momentum $\blk$ ejected by a probe pulse.  For an
arbitrary probe pulse of temporal profile $e(t)$ we have~\cite{perfetto2016first}
\eq{
 N_{\blk}(\w)=2\sum_{ij}\int dt dt'\;\Re\left[ \Sigma^{\rm R}_{ij\blk}\(\w,t,t'\)G^{<}_{ji\blk}\(t',t\)\right].
 \label{Nkw}
}
The ionization self-energy  reads~\cite{PUvLS.2015}
\eq{
  \Sigma^{\rm R}_{ij\blk}(\w,t,t')=-i\th(t-t')  d_{i\blk} e(t)e^{-i\w(t-t')} d_{j\blk} e(t'),
}
where $d_{i\blk}$ is the dipole matrix element between a state in band $i$ and momentum $\blk$ and a continuum time-reversed LEED state of energy  $\w$ and
parallel momentum $\blk$. 
\e{Nkw} is valid for systems in arbitrary non-equilibrium states and for any temporal shape and {\em intensity} of the probe field (hence even
beyond linear response), the only approximation being that LEED electrons do not interact with bound electrons.  A similar expression for $N_{\blk}(\w)$ was
derived in Ref.~\cite{FKP.2009}.
We notice also that the
photocurrent is a functional of the probe pulse $e(t)$ and, therefore, it depends on the time $T$ at which the pulse impinges the system.  However, from
\e{Nkw} we see that $N_{\blk}(\w)$ is a complicated two-times convolution of $G^{<}$.
 
A simpler and more transparent expression can be obtained if the probe pulse has a duration $t_{p}$ much longer than the typical electronic timescale and  a
frequency $\w_{p}$ large enough to resolve  the desired removal energies.  In this case one can shown that \e{Nkw} reduces
to~\cite{perfetto2016first,Perfetto2020a,Perfetto2020b,Perfetto2020d}
\eq{ 
 N_{\blk}(\w)\propto A^{<}_{\blk}(\t,\w-\w_{p}),
 \label{trarpessignal}
}
where $\t$ is the time at which the probe impinges the system and  $A^{<}_{\blk}$ is the transient (removal) spectral function 
\eq{
 A^{<}_{\blk}\(\t,\w\)=-i\int_{\t+\frac{t_{p}}2}^{\t-\frac{t_{p}}2}  d\bar{t}\,e^{i\w \bar{t}} \, \mathrm{Tr} 
 \[G^{<}_{\blk}\(\t+\frac{\bar{t}}2,\t-\frac{\bar{t}}2\)\].
 \label{trARPESAA}
}
In tr-ARPES experiments involving semiconductors with band-gap of the order of 1~eV, photo-excitation typically  occurs via a NIR/VIS pump pulses. As a
consequence we expect that the density matrix varies  on a timescale of few femtoseconds, given by the inverse of the lowest excitation energy.  Therefore the
approximated expression in Eq.~(\ref{trARPESAA}) can be safely used since typical probe fields are XUV pulses with duration $t_{p}$ of some tens of
femtoseconds.

\subsection{Coherent and incoherent regimes: preliminary remarks}
\lab{sec:coh}
As the title of this review highlights we want to discuss the role played by coherence in the TR--ARPES
phenomenology\cite{PhysRevB.100.115204,Madeo2020,Rustagi2018,Wallauer_2021,kemper2020observing,fukutani2021detecting}. As it was introduced in the
\sec{sec:motivations} by using simple arguments, the excitation with an external laser pulse transfers coherence in the system. The most striking evidence of
this coherence is the appearance of a finite polarization, which reflects the coherent oscillation of the system.

The finite polarization, however, is not the only fingerprint of a coherent dynamics. As it will be discussed in the following the microscopic properties of the
system will be affected and the a consequent rich phenomenology appears.

\begin{figure}[tbp]
    \centering
\includegraphics*[width=1\textwidth]{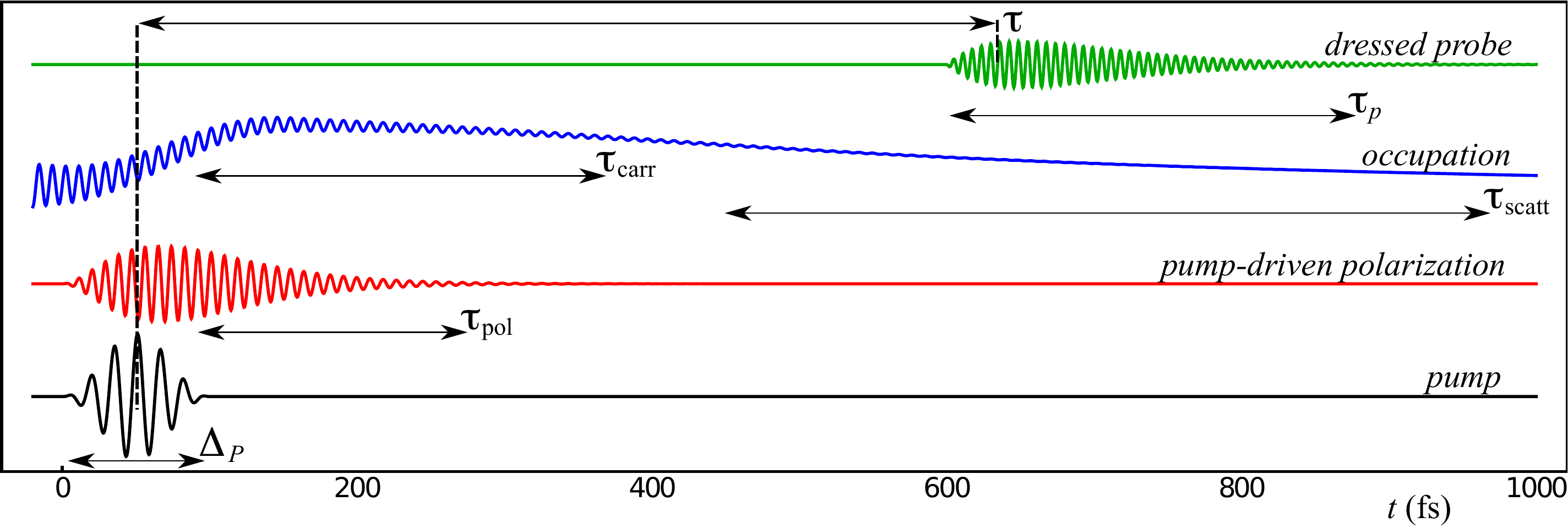}  
\caption{Illustration of the characteristic times occurring in a typical \ppe experiment, from \ocite{Perfetto2015}. $T_P$ is the time scale of the electron dynamics
induced by the pump, $\gt_{pol}$ is the dephasing time of the pump-induced polarization, $\gt_{carr}$ is the stabilization time of the occupations,
$\gt_{scatt}$ is the time to relax back to the equilibrium state, and $\gt_p$ is the lifetime of the dressed probe field. We also display the delay $\gt$ between pump and
probe}
\label{fig:coh}
\end{figure}

Let's, however, come shortly back on the definition of the coherent regime by using the arguments used in Figure 1 of \ocite{Perfetto2015}, that we reproduce
here in \fig{fig:coh}. After the pump field is applied it is well known that the system manifests a finite polarization that last for some time, $\gt_{pol}$.
Any experimentalist knows that this  time scale depends on the specific system. In extended materials can be as long as $\sim$100\,fs, while in finite systems
can be much longer.

While the pump pulse is active the quantum theory of elastic light scattering predicts the system to adsorb light at specific frequencies connected to the poles
of the macroscopic dielectric function, \e{eq:eps2}. These frequencies can be obtained by solving the BSE, \e{eq:BSE}. This is commonly done in Material Science
codes, like \yambo and for realistic materials it is found that the absorption frequencies can be located below the
optical gap or above.

In the first case we talk of {\em excitons}, bound electron--hole pairs, whose energy lays in the single--particle forbidden region. When the absorption peak is
located above the gap, instead, we talk of free electron--hole pairs. The physics of excitons is very rich and the microscopic properties of these exotic objects
has been extensively studied. While a comprehensive review here is impossible we would like to mention the excitonic states, like the electronic states, suffer
of scattering events. A paradigmatic example is the exciton--phonon scattering that in many 
works\cite{Cudazzo2020,Moody2015,Shree2018,Slobodeniuk2016a,Chen2020} has been proved to  broad the excitonic lines.

This broadening points to a internal structure of the excitonic states that reflect their balance between the scatterings of the individual electrons/holes that
participate to the excitonic state with the collective properties of the exciton as a new, bosonic like, particle. Actually the possibility to describe excitons
ad individual and well defined bosonic particles~\cite{Katsch2018} is an highly debated aspect that already count for a long history of
works~\cite{Combescot2017,moskalenko_snoke_2000}. 

Going back to \fig{fig:coh} we see that the coherent regime lives as long as the polarization lives. After the decay of the polarization, also named {\it
decoherence}, the system remains in an excited state where carriers are still populating the conduction bands. The question now is: are those carriers entirely
embodied in excitonic states? Alternatively we could ask how the coherent picture evolves in the incoherent regime.
As today there are no definitive answers to these questions. What we will review in the next sections is what we know on the coherent and incoherent regimes
{\em with arguments firmly based on the NEGF approach}.

To this end we will consider two systems in the two regimes:
(i) the two--bands model described by the $H^{eq}_{model}$, \e{minmodham}.
(ii) a selection of realistic materials studied by means of \ai\,NEGF. We will consider bulk Silicon and Black--Phosphorus.

\subsubsection{Excitonic signatures in the coherent regime of the model system}
\lab{sec:coh_excitons}
In the case of the Hamiltonian given by \e{minmodham}, the coherent regime, coincides with the 
{\it resonant pumping}, i.e. when the pump frequency $\omega_{P}$ matches the excitonic energy. 
In this case photo-excitation provides a direct and
efficient excitation of excitons, since quasi-particle states are accessible only for
$\omega_{P}>\epsilon_{g}$~\cite{Schmitt-Rink_PhysRevB.37.941,GLUTSCH1992,Hannewald-Bechstedt_2000,Christiansen2019,Perfetto2019a,Dendzik2020}.  
Indeed for $H^{eq}_{model}$ there exist only one exciton, which can be made bound by tuning the Hamiltonian parameters, with energy  $\epsilon_{X}$.

We first characterize the many--body of the two--bands model as an oscillatory state that is generated by shining the system with a pump field with  frequency $\omega_{P}=\epsilon_{X}$.  We
have recently shown~\cite{Perfetto2019a,Perfetto2020a,perfetto2020ultrafast} that after pumping the system is temporarily left in a BCS-like superfluid composed
by a macroscopic number of excitons having a finite and oscillating polarization.  Below we briefly discuss how to calculate the explicit wave-function of this
superfluid  since it is useful to predict the corresponding tr-ARPES spectrum.  

\begin{figure}[tbp]
    \centering
\includegraphics*[width=.8\textwidth]{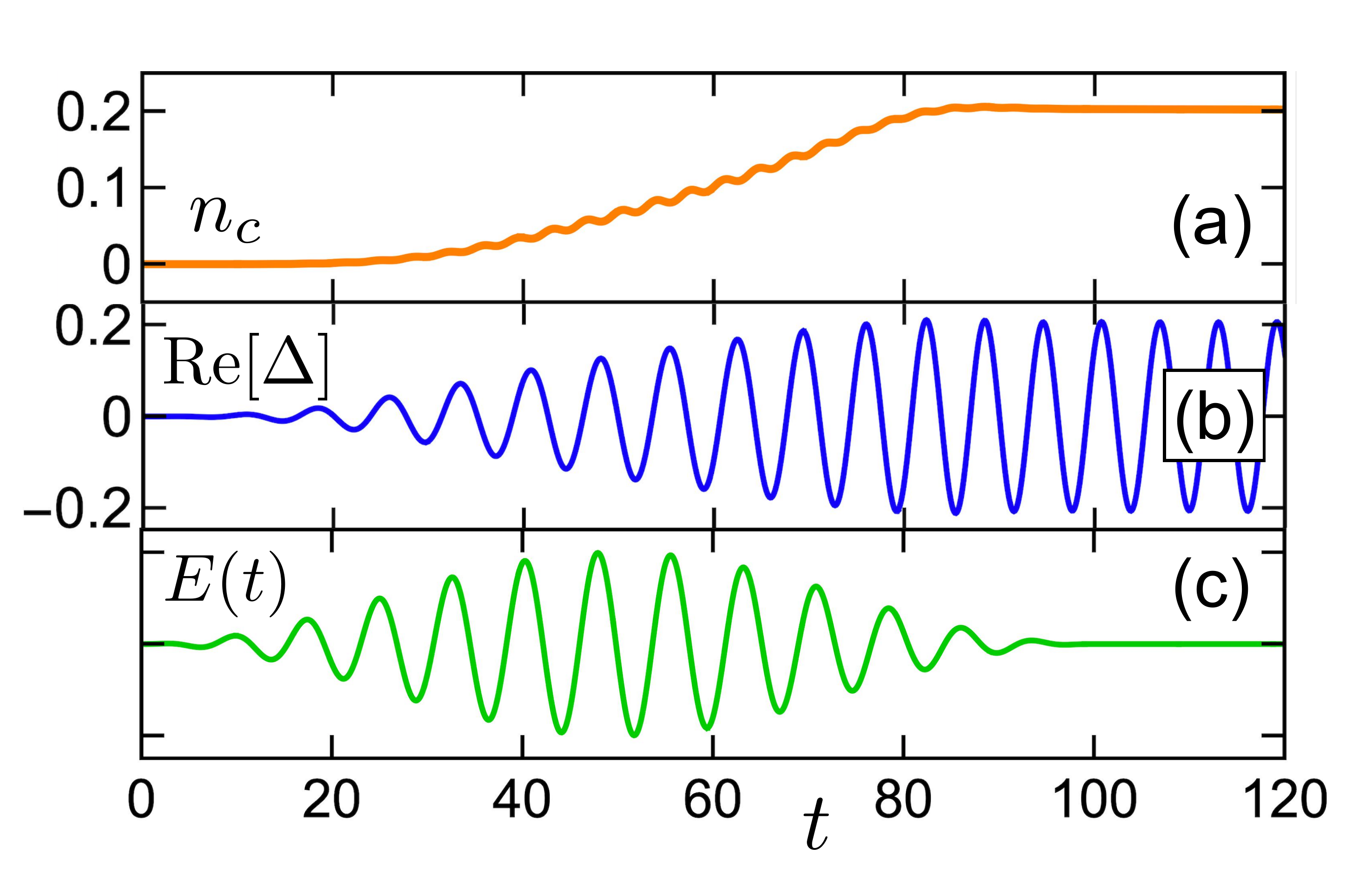}  
\caption{(a) Total density in the conduction band $n_{c}(t)$ and real part of the order parameter $\Delta(t)$ due to pumping with pump-profile shown in panel
(c).  We have considered the one-dimensional model in Eq.~(\ref{minmodham}) with bandwidth $w=2$, repulsion $U=1$, constant dipole moments $D_{k}=D$,  pumping
intensity $\Omega_{P}=ED=0.05$ (with $E$ the pump amplitude) and pumping frequency $\omega_{P}\approx \epsilon_{X}=0.76$.  Energies are in units of the band-gap
$\epsilon_{g}$ and times in units of $1/\epsilon_{g}$.}
\label{tdhf}
\end{figure}

We further introduce a pumping term in the two bands model as
\eq{
  \hat{V}(t)=E(t)\sum_{k } D_{k} \[\hat{c}^{\dag}_{k }\hat{v}_{k}+\hat{v}^{\dag}_{k}\hat{c}_{k}\].
 \lab{eq:V_mod}
}
\e{eq:V_mod} is added to $H^{eq}_{model}$ to mimic the effect of \e{eq:Vt} in \e{eq:full_h}.
The numerical results (see \fig{tdhf} and \ocite{Perfetto2019a}) show that after pumping there is a finite and constant occupation of the conduction
states $n_{ck}=\r_{cck}$, and a finite residual polarization $\D(t)=-\frac{1}{\mathcal{L}}\sum_{k}\r_{vc k}(t)$ that exhibits persistent monochromatic
oscillations at the excitonic frequency, i.e. $\D(t)=\D_{0}e^{i\epsilon_{X}t}$.

The crucial point here is that the oscillatory density matrix $\r(t)$ obtained from this simulation describes an excitonic superfluid, since it is {\it
numerically identical} to the density matrix obtained by solving a static and self-consistent BCS-like problem~\cite{Perfetto2019a}.  The peculiarity of this
mean-field problem consists in having different chemical potentials $\mu_{c}$ and $\mu_{v}$ for conduction and valence electrons, a pivotal feature to describe
an excited state.  The secular equation reads
\eq{
\left(
\begin{array}{cc}
    \epsilon_{vk} -\mu_{v} & U\D
    \\ U\D &  
    \epsilon_{vk} -\mu_{c} 
\end{array}
\right)  \left(
\begin{array}{c}
    \varphi^{\xi}_{vk}  \\ \varphi^{\xi}_{ck} 
\end{array}
\right) =  \varepsilon^{\xi}_{k} \left(
\begin{array}{c}
     \varphi^{\xi}_{vk}  \\ \varphi^{\xi}_{ck} \end{array}
\right) ,
\label{hk(z)}
}
where  $\xi=\pm$ labels the two eigensolutions, $\D=-\frac{1}{\mathcal{L}}\sum_{k} \varphi^{-}_{vk}\varphi^{-}_{ck}$ is the excitonic order parameter, and
$\varepsilon^{\xi}_{k}$ is the eigenvalue given by 
\eq{
 \varepsilon^{\xi}_{k}= \frac{\epsilon_{vk}+\epsilon_{ck} -\mu_{v}-\mu_{c} +\xi
 \sqrt{(\epsilon_{vk}-\epsilon_{ck} 
 -\mu_{v}+\mu_{c})^{2}+4U^{2}\D^{2}}  }{2}.
}
If the difference between the chemical potentials is larger than  the exciton energy $\epsilon_{X}$, i.e., $\delta \mu = \mu_{c} - \mu_{v} > \epsilon_{X}$, the
self-consistent problem in Eq.(\ref{hk(z)}) admits a non-vanishing order parameter $\D$, corresponding to an excitonic condensate
solution~\cite{Ostreich_1993,Hannewald-Bechstedt_2000,Yamaguchi_NJP2012,Yamaguchi_PhysRevLett.111.026404,Hanai2016,HanaiPRB2017,TriolaPRB2017,Hanai2018,PertsovaPRB2018,Perfetto2019a,Becker_PhysRevB.99.035304,Murakami2020}.
The superfluid is also characterized by a finite carrier density in conduction band 
\eq{
 n_{c}=\frac{1}{\mathcal{L}}\sum_{k}  |\varphi^{-}_{ck}|^{2}.
} 
The dilute limit $n_{c}<<1$ of the solution is very interesting.  In this case we have $\delta \mu \approx \epsilon_{X}$, and the order parameter $\D$ is very
small. Thanks to these simplifications the self-consistent solution of Eq.~(\ref{hk(z)}) can be analytically determined.  The lowest eigenvalue becomes
$\varepsilon^{-}_{k}=\epsilon_{vk}+\epsilon_{X}/2$, while the corresponding eigenvector $\varphi^{-}_{k}$ can be written in terms of the excitonic eigenvector
$Y_{k}$ of the BSE equation (which the form that the BSE, \e{eq:BSE} acquires for the two--bands model descried by \e{minmodham}):
\eq{
 \(\epsilon_{ck}-\epsilon_{vk}-\epsilon_{X}\)Y_{k}=\frac{U}{\mathcal{L}}
  \sum_{q}Y_{q}
 \lab{eq:bseminmod}
}
as
\seq{
\eqg{
 \varphi^{-}_{vk}\approx 1, \\  
 \varphi^{-}_{ck}\approx \sqrt{\mathcal{L}n_{c}}Y_{k}.
}}
Therefore the conduction component of the self-consistent solution is proportional to the exciton wave-function.  
 
From the solution of Eq.~(\ref{hk(z)}) we can readily calculate the superfluid Green's functions.  The Matsubara component  
$G^{\rm M}_{ij k}(\tau,\tau^{+})=i\varphi^{-}_{ik}\varphi^{-}_{jk} \equiv i \r^{\rm M}_{k ij}$ can be used  as initial condition to extract the lesser
component, having both arguments on the real axis:
\eq{
G^{<}_{ij  k}(t,t')=ie^{-i(\varepsilon^{-}_{k}-\sigma_{z}\frac{\delta\mu}{2})t}\r^{\rm M}_{k} e^{i(\varepsilon^{-}_{k}-\sigma_{z}\frac{\delta\mu}{2})t'} .
\label{gneq}
}
Notice that even if we have solved a static problem, $G^{<}(t,t')$ does {\it not} depend on the times difference.  As a consequence the density matrix
$\r_{k}(t)=-iG^{<}_{k}(t,t)$ is not-stationary, since it has constant occupations, but time-dependent off-diagonal elements:
\eq{
    \rho_{k}(t)=\left(
\begin{array}{cc}
    |\varphi^{-}_{vk}|^{2} & 
    \varphi^{-}_{vk}\varphi^{-}_{ck}e^{i\delta \mu t}
    \\  \varphi^{-}_{vk}\varphi^{-}_{ck} e^{-i\delta \mu t} &  
    |\varphi^{-}_{ck}|^{2}
\end{array}
\right) \equiv \left(
\begin{array}{cc}
    n_{vk} & 
    \Delta_{k}e^{i\delta \mu t}
    \\  \Delta_{k} e^{-i\delta \mu t} &  
    n_{ck}
\end{array}
\right).
\label{tdrho}
}
As anticipated, our TD-HF simulations show that after  weak resonant pumping  the density matrix of the system is numerically identical to the one of
Eq.~(\ref{tdrho}), obtained with $\delta \mu = \epsilon_{X}$.  This finding strongly suggests that an excitonic superfluid can be transiently created in real
time by pumping a normal insulator in resonance with a bright exciton.  Here the self-sustained oscillations generate a Floquet-like regime in the absence of
external driving~\cite{Ostreich_1993,SzymaPRL2006,Perfetto2019a}, and are expected to survive over a timescale dictated by the polarization lifetime.  We mention
that a particularly interesting instance may occur in systems with optically bright $p$-wave excitons.  In this case it has been recently shown that resonant
pumping could induce the creation of an exciton superfluid with nontrivial topological properties~\cite{Perfetto2020d}.

We have now all the ingredients to calculate the time-dependent photo-emission spectrum originating from the exciton superfluid.  By substituting the expression
for $G^{<}(t,t')$ in Eq.~(\ref{gneq}) in the formula (\ref{Nkw}) for the photocurrent, one can show that $N_{k}(\omega)$ takes the elegant and compact
form~\cite{Perfetto2020a}
\eq{
  N_{k}(\omega)=\left| \varphi^{- \dagger}_{k} \tilde{e}\left(
  \varepsilon^{-}_{k} \hat{\mathbb{1}}-\sigma_{x}\frac{\delta \mu}{2} -\omega \hat{\mathbb{1}} 
  \right) d_{k} \right|^{2},
  \label{nkwana}
}
where $\tilde{e}(\omega)$ is the Fourier transform of the probe pulse  $e(t)$.

\begin{figure}[tbp]
    \centering
\includegraphics*[width=1.0\textwidth]{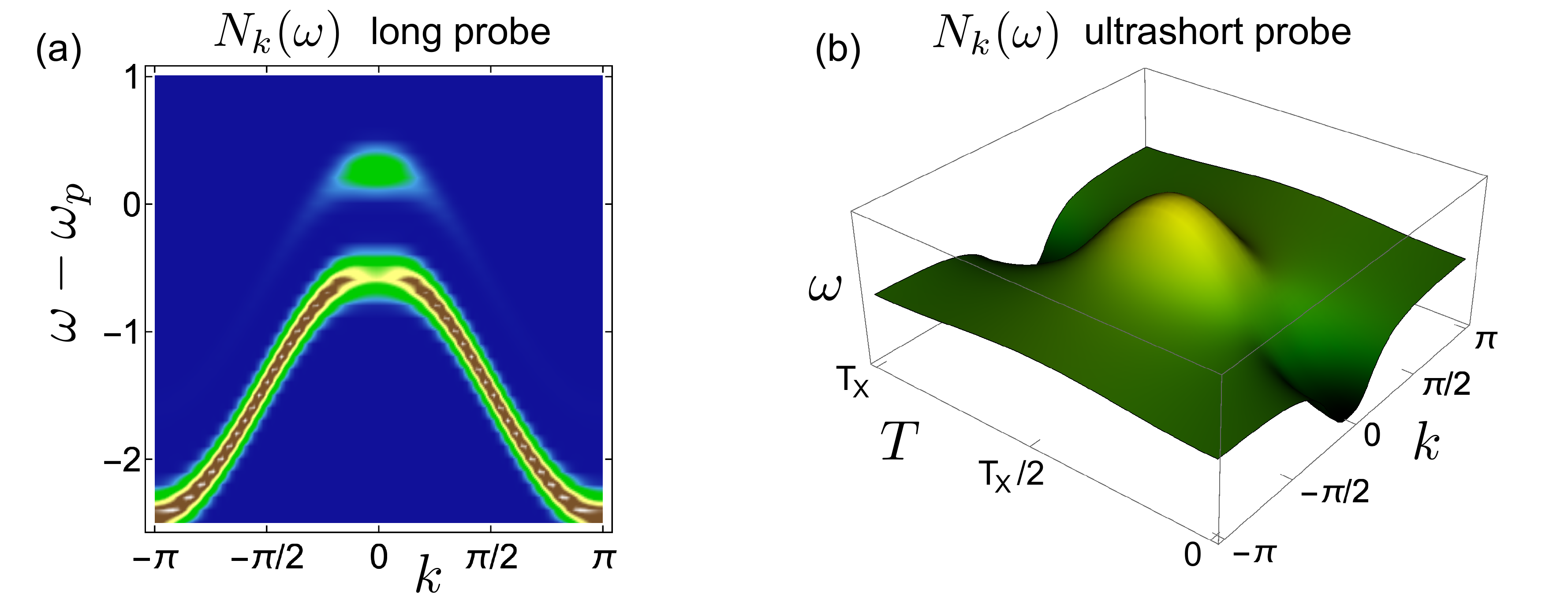}  
\caption{Transient photocurrent $N_{k}(\omega)$ for the one-dimensional system in Eq.~(\ref{minmodham}) for the same parameters as in \fig{tdhf} for
monochromatic (panel a) and  ultrashort (panel b) probe. Frequencies $\omega$ are measured with respect to the valence band maximum. Energies are in units of
the band-gap $\epsilon_{g}$ and times in units of $1/\epsilon_{g}$. } \label{alessmodel}
\end{figure}

It is instructive to further manipulate the above expression for low excited densities in two opposite cases, namely in the limit of a long--monochromatic
probe, and in the limit of ultrashort probe.  For $e(t)=e\sin[\omega_{p}(t-T)]$ ($T$ being the time at which the probe impinges the system) we have
\eq{
     N_{k}(\omega) \propto 
     \delta(\omega-\epsilon_{v}-\omega_{p})+\mathcal{L}n_{c}|Y_{k}|^{2} 
     \delta(\omega-\epsilon_{v}-\epsilon_{X}-\omega_{p}).
     \label{nkwlong}
}
The numerical evaluation of the above equation, for the same model parameters used in \fig{tdhf} is shown in \figlab{alessmodel}{a}.  We clearly observe the
occurrence of a quasi-particle peak at $\omega = \epsilon_{vk}$ corresponding to the removal an electron from the valence band, accompanied by an {\it excitonic
satellite} inside the gap, at $\omega = \epsilon_{vk}+\epsilon_{X}$. The latter corresponds to the removal of an electron from the excitonic superfluid. As a
function of the momentum, the exciton structure forms a side-band which is a replica of the valence band, shifted upward by
$\epsilon_{X}$~\cite{Perfetto2019a,Perfetto2020a}. The spectral weight of the side-band is proportional to the carrier density $n_{c}$, and to the square modulus
of the exciton wave-function $|Y_{k}|^{2}$. We also observe that in this case the photocurrent is essentially independent on the pump-probe delay $T$.

A comment on the physical origin of the excitonic side-band is in 
order.
In the coherent regime the density matrix has  off-diagonal elements that oscillate at frequency $\delta \mu$, see Eq.(\ref{tdrho}). Thus one can show that
gran-canonical HF Hamiltonian in Eq~(\ref{hk(z)}) coincides with the Floquet Hamiltonian of the same system, subject to monochromatic driving with frequency
$\delta \mu = \epsilon_{X}$.  According to the Floquet theory, a replica of the originally filled band  shifted upward by the driving frequency is expected to
occur in the  spectral function of the system~\cite{Perfetto2015}.   
 
We have shown that the above approach can be also implemented in ab-initio simulations of realistic materials.  To this end we have considered LiF
bulk~\cite{Perfetto2019a}, an  insulator with band-gap $\epsilon_{g}=14.5$~eV and having a $1s$ bright exciton with a binding energy $\epsilon_{X}=12.5$~eV, and
by using the \yambo code we have solved a first-principle version of the self-consistent problem in Eq.~(\ref{hk(z)}). Also in this case we have found a superfluid
solution for $\delta \mu \gtrsim \epsilon_{X} $, giving rise to a photocurrent spectrum with an excitonic replica of the valence band appearing inside the gap. 

The situation is completely different in the opposite limit of 
ultrashort probes. For $e(t)=e\delta(t-T)$ Eq.~(\ref{nkwana}) yields
\eq{
  N_{k}(\omega) \propto C_{k}+Y_{k}\cos(\varepsilon_{X}T+\phi_{k}),
    \label{nkwshort}
}
where $C_{k}$ is a $\omega$-independent  background and $\phi_{k}$ is a phase that depends on the dipole matrix elements $d_{k}$ and on the probe amplitude $e$.
In this limit  $N_{k}(\omega)$ is independent on $\omega$ and has nothing to do with the original band-structure of the system.  The photocurrent and is an
oscillating function of the delay $T$, with period $T_{X}=2\pi/\epsilon_{X}$ dictated by the exciton energy, see \figlab{alessmodel}{b}. The oscillatory term is
proportional to the exciton wave-function $Y_{k}$.  Thus it appears that tr-ARPES experiments in the coherent regime with long or ultrashort probes can be used
for a direct measurement of the exciton wave-function.  This has been very recently demonstrated in tr-ARPES experiments on transition metal dichalcogenides
employing probes with $T_{p}$ of several tens of femtoseconds~\cite{man2020experimental,dong2020measurement}.

\begin{figure}[tbp]
    \centering
\includegraphics*[width=1.0\textwidth]{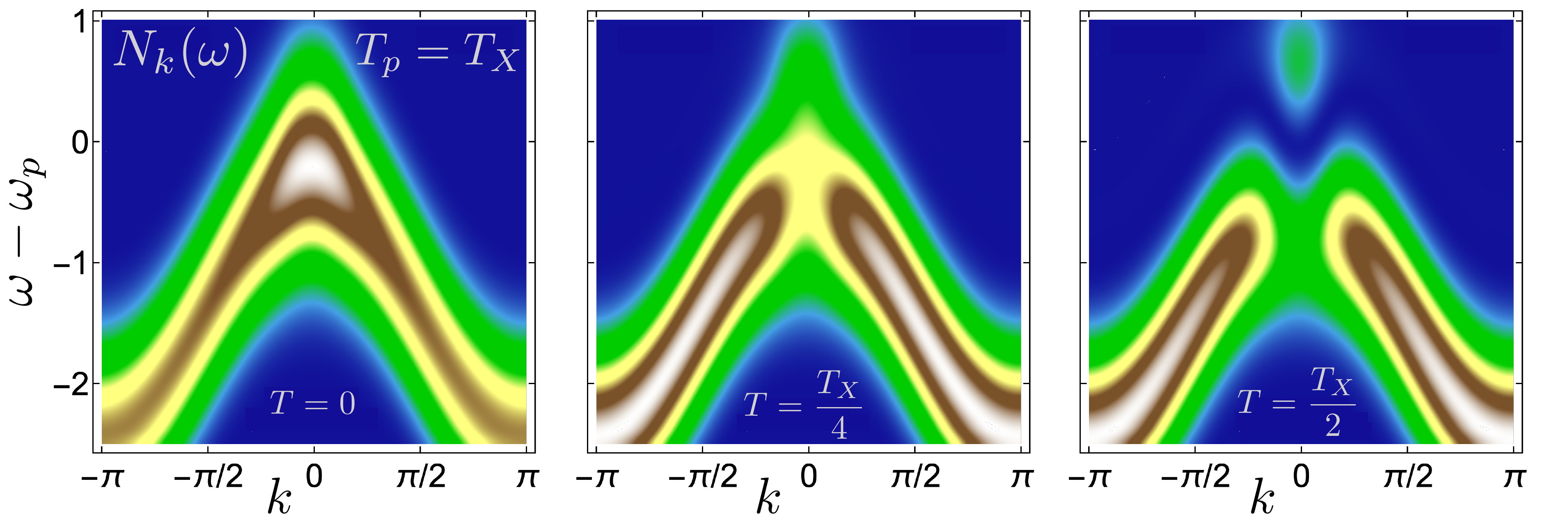}  
\caption{Transient photocurrent $N_{k}(\omega)$ for the one-dimensional system in Eq.~(\ref{minmodham}) for the same parameters as in \fig{tdhf} for a
sinusoidal probe of frequency $\omega_{p}=40$, duration $T_{p}=T_{X}\approx 8$ for three different delays $T\approx 0$, $T=T_{X}/4$, $T=T_{X}/2$. Frequencies
$\omega$ are measured with respect to the valence band maximum. Energies are in units of the band-gap $\epsilon_{g}$ and times in units of $1/\epsilon_{g}$.}
\label{nkwinter}
\end{figure}

In order to study the intermediate regime of finite-duration probes, we have to resume Eq.~(\ref{nkwana}).  In \fig{nkwinter} we display the transient
photocurrent for a sinusoidal probe of duration $T_{p}$ comparable with the excitonic period $T_{X}$.  We see that  $N_{k}(\omega)$ exhibits different features
depending on the delay $T$.  For small delays $T=0$, the tr-ARPES spectrum resembles the simple valence band. For $T=T_{X}/4$ excitonic effects start to be
visible, as there is a sizable transfer of spectral weight from the top of the valence band maximum toward higher energies.  For $T=T_{X}/2$, the probe impinges
the system when the real part of the order parameter is maximum, and the exciton satellite is fully developed inside the gap.  
   
We conclude this section by commenting on the robustness of the predictions made in this section.  The excitonic superfluid responsible for the peculiar
spectral features discussed above is characterized by a {\it finite} excited density $n_{c}$.  In principle excited carriers could screen the Coulomb
attraction, thus causing the disappearance of the superfluid phase.  However, in a recent paper we have shown~\cite{Perfetto2020a} that as long as the system
hosts a finite order parameter, the long-wave component of the dielectric function vanishes. Indeed excitons are neutral composite quasi-particles with very
scarce screening efficiency:  a gas of microscopic electric dipoles is not capable to modify the long-distance behavior of the Coulomb repulsion.  This
remarkable property renders the exciton condensate particularly robust, at least for small and moderate excited density.  

\subsection{Incoherent regime: excitons versus quasiparticles}
\lab{sec:incoh}
As discussed in \sec{sec:coh} the incoherent regime appears when the induced polarization is dead and there are still carriers populating the conduction bands.

At long times (of the order of tens of picoseconds) after pumping, it is well known that incoherent effects caused by carrier-carrier~\cite{Haug2004,steinhoff2016nonequilibrium} and
carrier--phonon~\cite{Selig2016,Molina-Sanchez2017} scattering drive the system towards a relaxed and dephased regime where electrons (holes) remain trapped in
the conduction (valence) band and place around the conduction band minimum (valence band maximum).  

If we consider systems characterized by a sizable electron--hole attraction it is clear, from simple arguments, that the excited carriers will form bound
states. At the same time it is natural to expect that these bound states will coexist with the plasma of free carriers in an excited, quasi--stationary state 
whose description is a real challenge for theoreticians~\cite{ulbricht2011carrier,KKKG.2006,perfetto2016first,Steinhoff2017}.

In the following two sections we will tackle the problem from two very different perspectives: again using the two--bands model, \e{minmodham}, to demonstrate how
the theory can capture the binding of the electron--hole photo--excited pairs; and the carriers perspective where we describe, using \ai\, methods the carrier
dynamics in realistic materials and directly comparing to the experimental results.

\subsubsection{Incoherent excitonic contribution in the model system}
\lab{sec:incoh_excitons}
We consider again the two--bands model, \e{minmodham}, and we 
we start observing that in this incoherent regime $G^{<}$ is diagonal in the band-index and depends only on the times-difference 
\eq{
 G_{ij \blk}^{<}\(t,t'\)\approx \delta_{ij}G_{ii \blk}^{<}\(t-t'\).
}
Therefore the transient spectral function in  Eq.~(\ref{trARPESAA}) is weakly dependent on the time $T$ at which the probe arrives, and can be approximated as
\eq{
 A^{<}_{\blk}(\t,\w)\approx A^{<}_{\blk}(\w)\approx -i \sum_{i}  G^{<}_{ii \blk}(\omega).
 \label{specinc}
}
In addition, experiments~\cite{CRCZGBBKGP.2012,SYACFKS.2012,NOHFMEAABEC.2014,Bernardi2014} and numerical simulations~\cite{Sangalli2015a,Sangalli2016}, indicate
that the electron occupations in the quasi-stationary excited state follow a Fermi-Dirac distribution with temperatures $T_{i}$ and chemical potentials $\m_{i}$
depending on the band index $i$.  Clearly both $T_{i}$ and  $\m_{i}$ vary on a picosecond time-scale but they can be considered as constant on the time-scale of
the probe pulse. From such experimental and numerical evidence we infer  that the recombination of electrons with different band index is severely suppressed
and hence  the lesser Green's function fulfills the approximate fluctuation-dissipation relation
\eq{
 G^{<}_{ii \blk}\(\w\)=-f_{i}\(\w\)\[G^{\rm R}_{ii \blk}(\w)-G^{\rm  A}_{ii \blk}(\w)\],
\label{heuristic6}
}
where $f_{i}(\w)=1/(e^{(\w-\m_{i})/T_{i}}+1)$ is the Fermi function for the  band $i$. The $i$-dependent temperature and chemical potential can be extracted by
a best fitting of the electronic populations as obtained from, e.g., the one-time GKBA propagation as it has been shown in Ref.~\cite{Sangalli2015a}.  From
Eq.~(\ref{heuristic6}) it appears that we are left with the evaluation of the retarded Green's function
\eq{
G^{\rm R}_{ii \blk}(\w)=\left[\frac{1}{\w-h^{\rm  qp}_{\blk}-\gS^{\rm R}_{\blk}(\w)} \right]_{ii},
\label{graexact}
}
where the $T$-matrix self-energy $\Sigma^{\rm R}_{\blk}$ carries all the information about the excitonic spectral features.

We now assume that the system is in a quasi-stationary excited state with a finite population in the conduction band, and show how to construct explicitly the
$T$-matrix self-energy accounting for exciton signatures in the spectral function.  We focus on the ARPES signal belonging to conduction electrons, and
therefore in \e{specinc} we evaluate only the contribution $A_{cc k}(\omega) \equiv -i G^{<}_{cc \blk}(\omega)$. To this end we need the
conduction-conduction self-energy $\gS_{cc p}(\w)$.

In the relaxed excited state, let $f_{i p}=\r_{ii p}$ be the population of the state in band $i=v,c$ with momentum $p$.  Accordingly, the undressed
quasi-particle Green's functions in the HF approximation read
\seq{
\label{gundr}
\eqg{
g^{<}_{i p}(\w)=2\p i f_{i p}\delta(\w-\gee_{i p}),\\
g^{>}_{i p}(\w)=-2\p i \bar{f}_{i p}\delta(\w-\gee_{i p}),
}
}
with $\bar{f}_{i p}=1-f_{i p}$. Using these quasi-stationary Green's function to calculate the lesser/greater $T$-matrix self-energy we find
\seq{
\label{selfT}
\eqg{
\gS^{<}_{cc p}(\w) =iU^{2}\sum_{q}f_{v p-q} L^{q,<}(\w-\gee_{v 
p-q}),\\ 
\gS^{>}_{cc p}(\w) = -iU^{2}\sum_{q}\bar{f}_{v p-q}L^{q,>}(\w-\gee_{v p-q}),
}}
where $L^{\lessgtr}$ are the lesser/greater components of the dresses electron-hole propagator. They are related to the retarded and advanced components
$L^{\rm R/A}$ via the relations~\cite{PSMS.2016}
\seq{
\eqg{
L^{q,<}=-2\eta \sum_{kk'p}  L^{q,\rm R}_{kp}(\w) \frac{f_{cp+q}\bar{f}_{vp}}{(f_{vp}-f_{cp+q})^{2}} L^{q,\rm A}_{pk'}(\w), \label{L<approx2}\\
L^{q,<}=-2\eta \sum_{kk'p}  L^{q,\rm R}_{kp}(\w) \frac{\bar{f}_{cp+q}f_{vp}}{(f_{vp}-f_{cp+q})^{2}} L^{q,\rm A}_{pk'}(\w). \label{L>approx2}
}}
The equations close as the retarded electron-hole propagator satisfies the BSE, \e{eq:BSE}
\eq{
 L^{q,\rm R}_{kk'}(\w)= L_{0,kk'}^{q,\rm R}(\w)+
 \frac{i}{\mathcal{L}}\sum_{p}L_{0,kp}^{q,\rm R}\,U\,L^{q,\rm R}_{pk'}(\w)
 \label{bse}
}
where non-interacting propagator is evaluated with the undressed 
Green's functions in Eq.~(\ref{gundr}) and  is given by
\eq{
L_{0,kk'}^{q,\rm R}(\w)=i\delta_{kk'}\frac{f_{vk}-f_{c k+q}}{\w-\gee_{c k+q}+\gee_{vk}+i\h}.
}
Having the lesser/greater self-energy we can calculate $\gS^{\rm R}_{cc}$ from the Hilbert transform of $\gS^{>}_{cc}-\gS^{<}_{cc}$.  The retarded Green's
function therefore is calculated as
\eq{
 G^{(\rm R)}_{cc k}(\omega)=\frac{1}{\omega-\epsilon_{ck}-\gS^{\rm  R}_{cc k}(\omega)},
\label{grcc}
}
and according to  \e{heuristic6}, the desired conduction spectral function is 
\eq{
\label{acc}
A_{cc k}(\omega)=-2f_{c}(\omega)\mathrm{Im}[G^{(\rm R)}_{cc k}(\omega)].    
}
\begin{figure}[tbp]
    \centering
\includegraphics*[width=.6\textwidth]{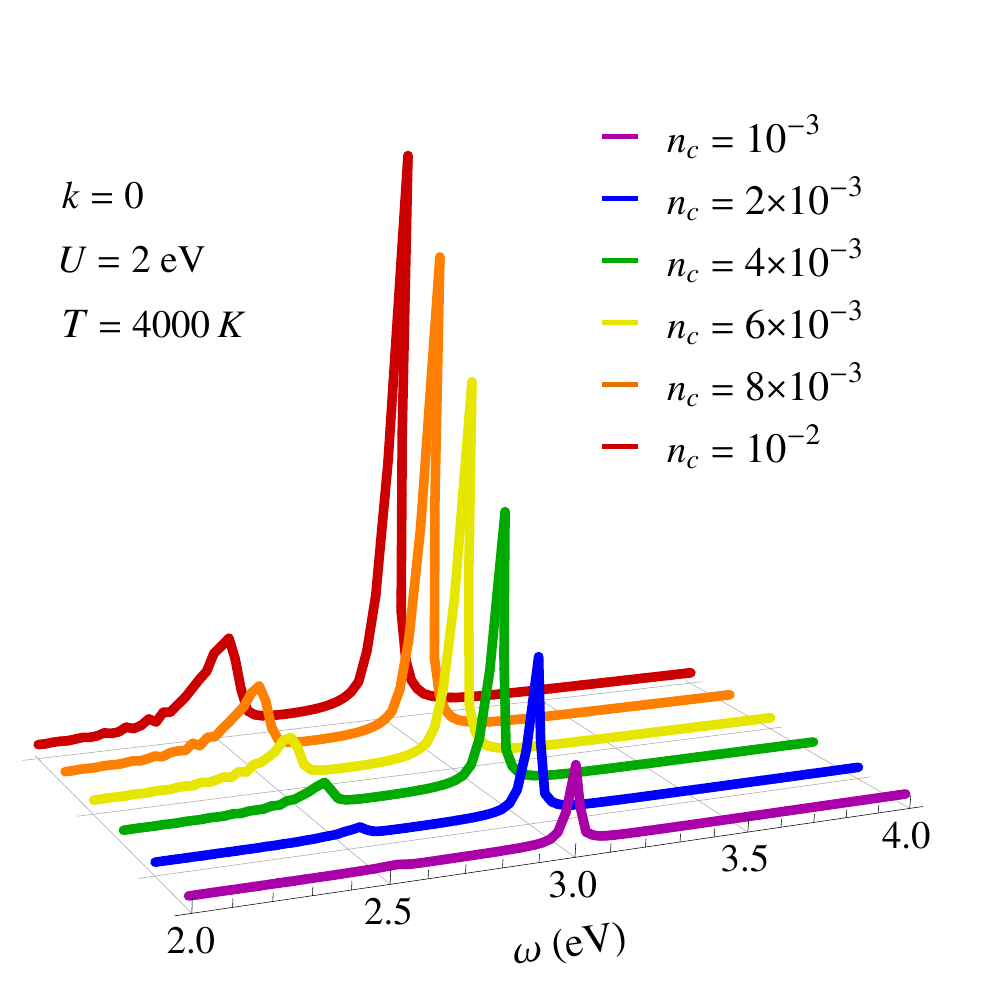}  
\includegraphics*[width=.3\textwidth]{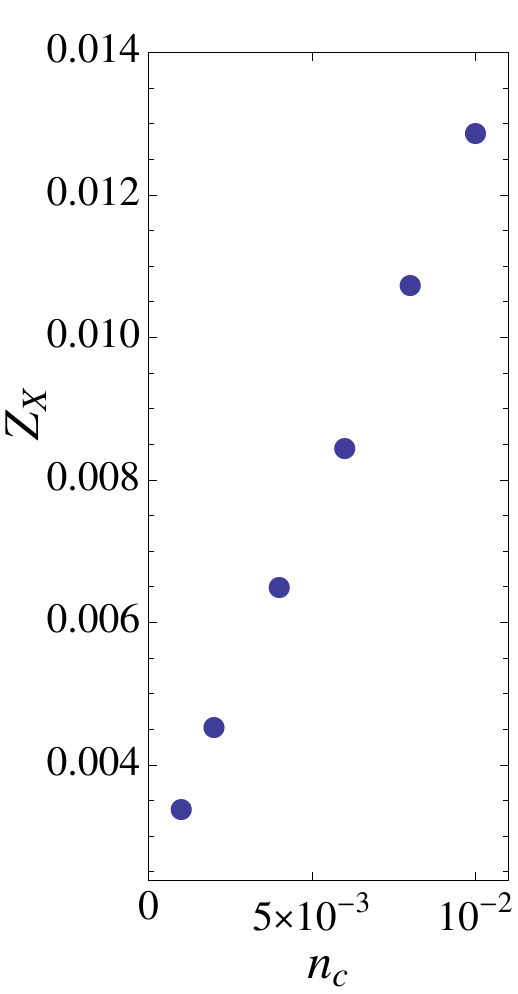}  
\caption{Left panel: Spectral function $A_{cc , 0}(\w)$ (in arbitrary units) for different densities of the conduction electrons $n_{c}$ at temperature
$T=4000$~K. Right panel: Dependence of the exciton weight $Z_{X}$ on $n_{c}$.  The convergence parameters are $\eta=w/(4\mathcal{L})$ and $\callL=80$.}
\label{specD}
\end{figure}
We observe that the vertex correction in the self-energy carried by the electron-hole correlator $L$ is crucial to produce the excitonic satellite in the
spectral function. Indeed the non-perturbative summation in the BSE, \e{bse}, generates the exciton peak in $L^{q,\rm R}(\w)$, occurring at energy
$\epsilon_{X} = \epsilon_{c0}-\epsilon_{v0}-b_{X}$, i.e {\it below} the onset of the particle-hole continuum. The presence of such split-off structure is then
transmitted to the self-energy via \e{selfT}.  

Explicit calculations have been performed by considering a model dispersion $\gee_{vk}=(w/2)\!\cos k$
and $\gee_{ck}=-(w/2)\!\cos k +w +\epsilon_{g}$, with bandwidth $w=4$~eV, band-gap $\epsilon_{g}=1$~eV, and repulsion $U=2$~eV. With these parameters the
conduction band minimum occurs at $\epsilon_{c0}= 3$~eV, and there is an exciton state with binding energy $b_{X} \approx 0.42$~eV.  The excited distribution
functions are taken at temperatures $T=T_{v}=T_{c}=4000$~K with $\mu_{v}=2.35$~eV and $\mu_{c}=2.65$~eV.  In Fig.~\ref{specD} (left panel) we show $A_{cc
0}(\w)$ for different carrier densities $n_{c}=\frac{1}{\mathcal{L}}\sum_{k}f_{ck}$.  At very low density $n_{c}\lesssim 10^{-4}$ the system is essentially in
equilibrium and the photocurrent is vanishingly small   (not shown).  At density  $n_{c} \approx 10^{-3}$ a quasi-particle peak at $\w \approx 3\, \mathrm{eV}$
appears.  This corresponds to the removal energy of an excited electron from the bottom of the conduction band.  As the density increases, a satellite appears
inside the gap, at energy $\gee_{c0}-b_{X} \approx 2.5\, \mathrm{eV}$.  This is clearly the exciton fingerprint discussed above. 
\begin{figure}[tbp]
  \centering
\includegraphics*[width=.8\textwidth]{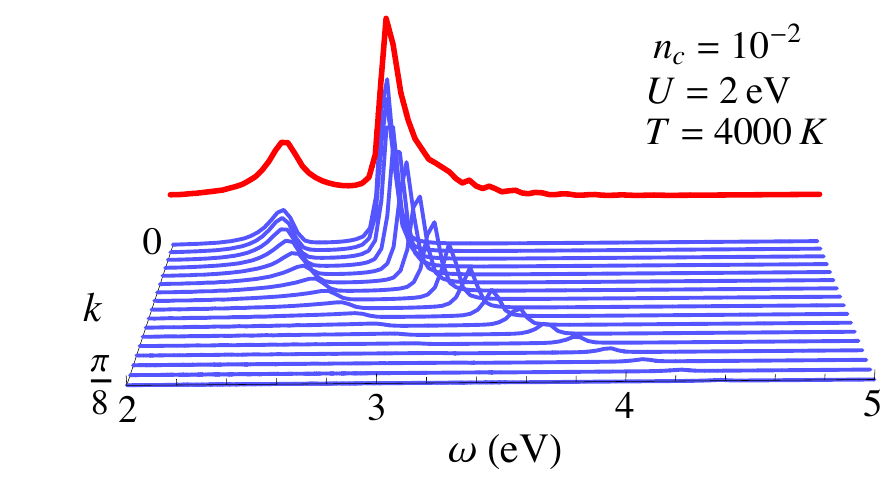}  
\caption{Spectral function $A_{cc  k}(\w)$ (in arbitrary units) for different momenta $k$. The (red) curve in the background is the integrated quantity $\int dk
\,A_{cc , k}(\w)$.  Same parameters as in Fig.~\ref{specD} and $n_{c}=10^{-2}$.}
\label{specK}
\end{figure}
We can also calculate the spectral function for different momenta $k$ of the conduction electron. This quantity is relevant to address angle-resolved
experiments.  In Fig.~\ref{specK} we plot $A_{cc  k}(\w)$ in the range $0<k<\p/8$.  We observe that the angle-resolved photocurrent displays and {\it excitonic
sideband} inside the gap, which is almost parallel to the conduction band. Similar results have been recently found by other authors by using an analogous
$T$-matrix approach to describe highly excited transition metal dichalcogenides~\cite{Steinhoff2017}.

\subsubsection{Coherent excitation and incoherent carriers migration}
\lab{sec:incoh_carriers}
A crucial result of the previous section is the possibility, in the model, to calculate the 
excitonic weight 
\eq{
 Z_{X} = -\int_{-\infty}^{\gee_{c0}} \frac{d\w }{2\p}\Im\[G^{\rm R}_{cc,0}\(\w\)\].
 \lab{eq:ZX}
} 
It has been pointed out that this quantity provides an estimate of the number of incoherent excitons coexisting with the plasma of free carriers in  conduction band~\cite{Steinhoff2017}.  From the right
panel of  Fig.~\ref{specD} we see that $Z_{X}$ increases with the density following a linear trend. 
From this analysis is evident that free carriers and bound states will coexist and, as it is numerically still unfeasible to take into account, within an \ai\,
scheme, excitons in the out--of--equilibrium dynamics the question is to what extent a purely carriers description is adequate?

Most of the methods currently used in the Material Science community are not based on formal justification, as it is not possible to derive analytical solution
for realistic materials. The methods are based on a Test\,\&Validate scheme. Approximations, like DFT--LDA, are applied on a many materials and the practical
justification of the approach is revealed from the success in describing the experimental results. In practice also the $GW$/BSE scheme has been devised in this
way~\cite{Onida2002}.

In order to formally move from the excitonic picture\,(introduced in the model) to the carriers representation we start by approximating the 
transient spectral function in \e{trARPESAA}. We  assume that the system is instantaneously in a
{\it quasi-equilibrium} state characterized by well--defined quasiparticles and by a vanishing polarization.  Under this adiabatic condition, we can use the GKBA
to express the quasi-particle $G^{<}$ in \e{trARPESAA} as
\ml{
 G^{<}_{\kk ij}\(T+\frac{\bar{t}}{2},T-\frac{\bar{t}}{2}\) \approx\\
 -\delta_{ij}\r_{\blk ii}\(T\)\[G^{\rm  R}_{\blk ii}\(\t+\frac{\bar{t}}{2},\t-\frac{\bar{t}}{2}\)-G^{\rm  A}_{\blk
  ii}\(\t+\frac{\bar{t}}{2},\t-\frac{\bar{t}}{2}\)\] \\ 
 \approx -\delta_{ij}\r_{\blk ii}\(T\)\[G^{\rm R}_{\blk ii}\(\bar{t}\)-G^{\rm A}_{\blk ii}\(\bar{t}\)\],
 \label{g<adiab}
}
where in the last line we have used the quasi-equilibrium assumption  that  $G^{\rm R/A}$ depends only on the time difference $\bar{t}$.  Therefore for
long-probe duration $t_{p}$ the transient spectral function becomes
\eq{
 A^{<}_{\blk}\(\t,\w\)\approx i \sum_{i} \r_{\blk ii}\(T\) \[G^{\rm  R}_{\blk ii}\(\omega\)-G^{\rm A}_{\blk ii}\(\omega\)\], 
 \label{trAadiab}
}
where 
\eq{
 G^{\rm  R/A}_{\blk ii}\(\omega\) =\left[\frac{1}{\omega -\uu{h}_{\blk}^{\rm qp} \pm i\eta}\right]_{ii} \approx \frac{1}{\omega -\epsilon_{i \blk}^{\rm qp} \pm 
 i\eta}.
 \label{grfree}
}
We thus arrive at the result that the tr-ARPES spectrum is proportional to the sum of the equilibrium spectral functions of the different bands $i$, weighted
by the corresponding time-evolving occupations $f_{\kk i}\(T\)=\r_{\blk ii}(T)$.   

If this approach does not allow to introduce incoherent excitonic effects in sense of \sec{sec:incoh_excitons} it allows to introduce
de--coherence effects in the form of electron--electron and electron--phonon scattering. These processes, as it will clear shortly, are essential to
quantitatively capture the physics observed experimentally in realistic materials.

The scheme that we will describe in the following is currently implemented in the \yambo\, code and it has been applied on a wealth of systems and observable:
it has been applied to 2D Semiconductors to calculate transient absorption\cite{Smejkal2021,Trovatello2020,Pogna2016}, Kerr
angle\cite{Wang2018,Molina-Sanchez2017}, TR--ARPES~\cite{Roth2019,Sangalli2015}.

In {\it Yambo} the out--of--equilibrium dynamics is split in two parts: the coherent excitation and the carriers scattering. In practice this means that the
KBE, \e{kbe}, is reduced, by using the GKBA and the MA~\cite{Marini2013,Melo2016} to a non--linear differential equation for $\r_{\kk ii}\(T\)$:
\eq{
 \frac{\partial}{\partial T} f_{\kk i}\(T\)= \left.\frac{\partial}{\partial T} f_{\kk i}\(T\)\right|_{coh}+\left.\frac{\partial}{\partial T} f_{\kk i}\(T\)\right|_{scatt},
\label{eq:carr.1}
}
with the scattering term composed of an electron--electron and electron--phonon contribution:
\eq{
\left.\frac{\partial}{\partial T} f_{\kk i}\(T\)\right|_{scatt}=
\left.\frac{\partial}{\partial T} f_{\kk i}\(T\)\right|^{e-e}_{scatt}+
\left.\frac{\partial}{\partial T} f_{\kk i}\(T\)\right|^{e-p}_{scatt}.
\label{eq:carr.2}
}

{\em The coherent excitation.}
The coherent term in \e{eq:carr.1}, as encoded in {\it Yambo}, fully takes into account excitonic effects. In practice this means that \e{eq:drhodt} is used with
the HF self--energy replaced by the statically screened $GW$ self--energy\,(SEX approximation). More details can be found in \ocite{Attaccalite2011b}. In practice we have that
\eq{
\left.\frac{\partial}{\partial T} \uu{\r}_{\kk}\(T\)\right|_{coh}=
-\[ \uu{h}^{KS}_{\kk}+\uu{\gd V}^{Hartree}_{\kk}\(T\)+\uu{\gd \gS}^{SEX}_{\kk}\(T\)+ \uu{V}_{\kk}\(T\),\r_{\kk}\(T\) \],
\lab{eq:carr.3}
}
with  $\gd V^{Hartree}$ and $\gd \gS^{SEX}$ the variations with respect to equilibrium of the Hartree and SEX self--energies. \e{eq:carr.3} gives is exactly
equivalent to the static BSE commonly applied in modern \ai\, codes\cite{Onida2002}. The coherent scattering term in \e{eq:carr.1} ensures that the excitation
process correctly capture excitonic effects and quantitatively reproduces the experimental excitation process.

{\em The scattering processes} 
are embodied in the second term of \e{eq:carr.1} and are the result of the application of the Markovian approximation, \sec{sec:MA}, and GKBA, 
\sec{sec:GKBA}, to the $GW$ self--energies described in \sec{sec:MA_self_energies}. The resulting e--e and e--p scattering terms are:
\ml{
\left.\frac{\partial}{\partial T}  f_{\kk i}\(T\)\right|_{scatt}^{e-p}=
-    \gc_{\kk i}^{\(e-p,\nearrow\)}\(T\)f_{\kk i}\(T\)
+\oo{\gc}_{\kk i}^{\(e-p,\nearrow\)}\(T\)\oo{f}_{\kk i}\(T\)\\
-    \gc_{\kk i}^{\(e-p,\swarrow\)}\(T\)f_{\kk i}\(T\)
+\oo{\gc}_{\kk i}^{\(e-p,\swarrow\)}\(T\)\oo{f}_{\kk i}\(T\)
\label{eq:dfdt_ep}
}
In \e{eq:dfdt_ep} the $\gc_{\kk i}^{\(e-p,\frac{\nearrow}{\swarrow}\)}$ are the NEQ carrier lifetimes which describes the scattering of carriers from one level and momenta  to
another by absorbing($\swarrow$)/emitting($\nearrow$) a phonon.  
\seq{
\label{eq:gamma_ep}
\eqg{
\gc_{\kk i}^{\(e-p,\frac{\nearrow}{\swarrow}\)}\(T\)= \pi \sum_{j \qq \mu} \frac{\lb g^{\qq\mu}_{\kk i j} \rb^2 }{N_\qq} \gd\(\gee_{\kk-\qq j}-\gee_{\kk i}\pm \go_{\qq \mu}\) 
\begin{pmatrix}
 1+f_{Bose}^{\qq\mu}\(T\) \\
 f_{Bose}^{\qq\mu}\(T\)
\end{pmatrix}
\oo{f}_{\kk-\qq j}\(T\),\\
\oo{\gc}_{\kk i}^{\(e-p,\frac{\swarrow}{\nearrow}\)}\(T\)= \sum_{j \qq \mu} \frac{\lb g^{\qq\mu}_{\kk i j} \rb^2 }{N_\qq} \gd\(\gee_{\kk-\qq j}-\gee_{\kk i}\pm \go_{\qq \mu}\) 
\begin{pmatrix}
 f_{Bose}^{\qq\mu}\(T\) \\
 1+f_{Bose}^{\qq\mu}\(T\)
\end{pmatrix}
f_{\kk-\qq j}\(T\).
}
}
In \e{eq:dfdt_ep} and \e{eq:gamma_ep} we have labeled the hole counterparts of the occupations and line-widths as $\oo{f}=2-f$ and $\oo{\gc}$. Moreover
$f_{Bose}^{\qq\mu}\(T\)$ is the occupation of the phonon mode $\qq\mu$ at temperature $T$. Thus \e{eq:gamma_ep} naturally embody the temperature dependence.

\begin{figure}[H]
\centering
\includegraphics[width=0.8\textwidth]{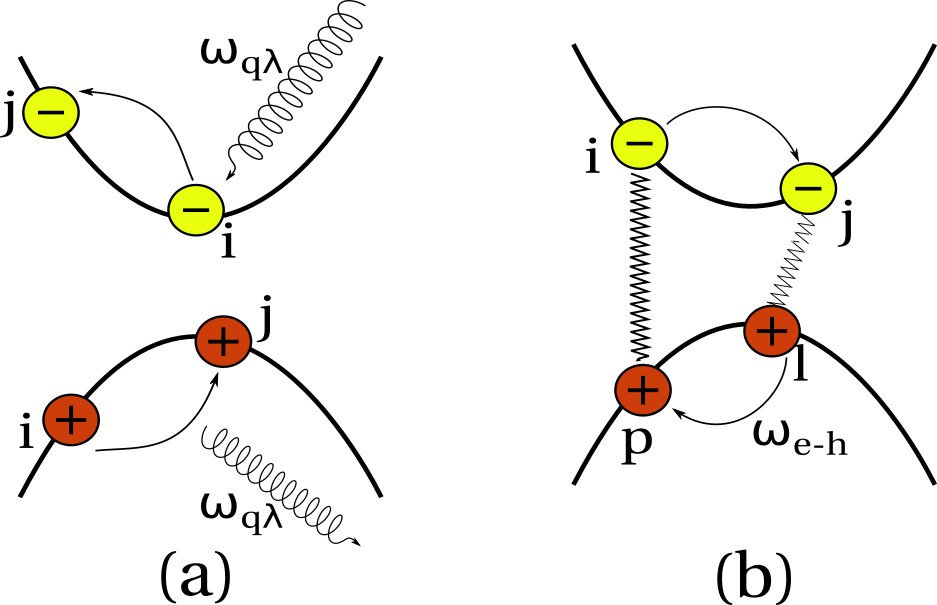}
\caption{Schematic representation of the elemental carrier scattering processes mediated by phonons and electron--hole pairs. From Ref.~\cite{Melo2016}.
The process\,(a) represents the physics of $\gc_{\kk i}^{\(e-p,\swarrow\)}\(T\)$ in the conduction and  $\oo{\gc}_{\kk i}^{\(e-p,\nearrow\)}\(T\)$ in the
valence, \e{eq:dfdt_ep}. In the (b) case, instead, the scattering $\(\kk i\)\rar \(\kk-\qq\)$ is energetically compensated by the promotion of an electron in
the valence. This is $\gc_{\kk i}^{\(e-e\)}\(T\)$, from \e{eq:gamma_ee}. }
\label{fig:carrier_scatts}
\end{figure}

A similar derivation can be done for the e--e channel~\cite{Marini2013,Melo2016}:
\eq{
\left.\frac{\partial}{\partial T}  f_{\kk i}\(T\)\right|_{scatt}^{e-e}=
     \gc_{\kk i}^{\(e-e\)}\(T\)f_{\kk i}\(T\)
-\oo{\gc}_{\kk i}^{\(e-e\)}\(T\)\oo{f}_{\kk i}\(T\),
\label{eq:dfdt_ee}
}
with
\seq{
\label{eq:gamma_ee}
\ml{
\gc_{\kk i}^{\(e-e\)}\(T\)=  \frac{4\pi}{N_\qq\Omega} \sum_{\qq \pp }\sum_{jlm} \lb W^{\qq}_{\substack{i j \kk \\ m l \pp }} \rb^2 \\
 \gd\(\gee_{\kk-\qq j}+\gee_{\pp m}-\gee_{\pp-\qq l}-\gee_{\kk i}\) \oo{f}_{\kk-\qq j}\(T\) \oo{f}_{\pp m}\(T\)f_{\pp-\qq l}\(T\),
}
and
\ml{
\oo{\gc}_{\kk i}^{\(e-e\)}\(T\)=  \frac{4\pi}{N_\qq\Omega} \sum_{\qq \pp }\sum_{jlm} \lb W^{\qq}_{\substack{i j \kk \\ m l \pp }} \rb^2 \\
 \gd\(\gee_{\kk-\qq j}+\gee_{\pp m}-\gee_{\pp-\qq l}-\gee_{\kk i}\) f_{\kk-\qq j}\(T\) f_{\pp m}\(T\)\oo{f}_{\pp-\qq l}\(T\),
}
}
\e{eq:gamma_ep} and \e{eq:gamma_ee} describes a simple $\(\kk i\)\rar \(\kk-\qq\)$ mediated by phonons and electron--hole pairs. The scatterings can have
different kinematic geometries and few of them are schematically discussed in \fig{fig:carrier_scatts}. In the left frame an electron  excited in the
conduction bands, in the state $i$ scatters to an higher state , $j$ by absorbing a phonon $\go_{\qq\mu}$. In the same frame a hole in the state $i$ is filled
by an electron from state $j$ by emitting a phonon. These two processes alter the phonon population.

Similarly, in the right frame of  \fig{fig:carrier_scatts}, the electron $i$ moves to a lower state, $j$ by releasing the energy which is used to promote the
electron in the $p$ state to the final, higher, $l$ state.

\subsection{Time--Resolved ARPES from first--principles in two paradigmatic 
            materials: bulk Silicon and Black--Phosphorus}
\lab{sec:materials}
\e{eq:carr.1} has been applied in
several materials and to describe a wealth of time--dependent physical observable.
The specific case of the TR--ARPES was the subject of \ocite{Roth2019} and \ocite{Sangalli2015} where two paradigmatic materials are investigated: bulk Silicon
and bulk Black--Phosphorus.

\subsubsection{Symmetry breaking and ultra--fast restoring in Bulk Silicon}
\begin{figure}[H]
\centering
\includegraphics[width=0.6\textwidth]{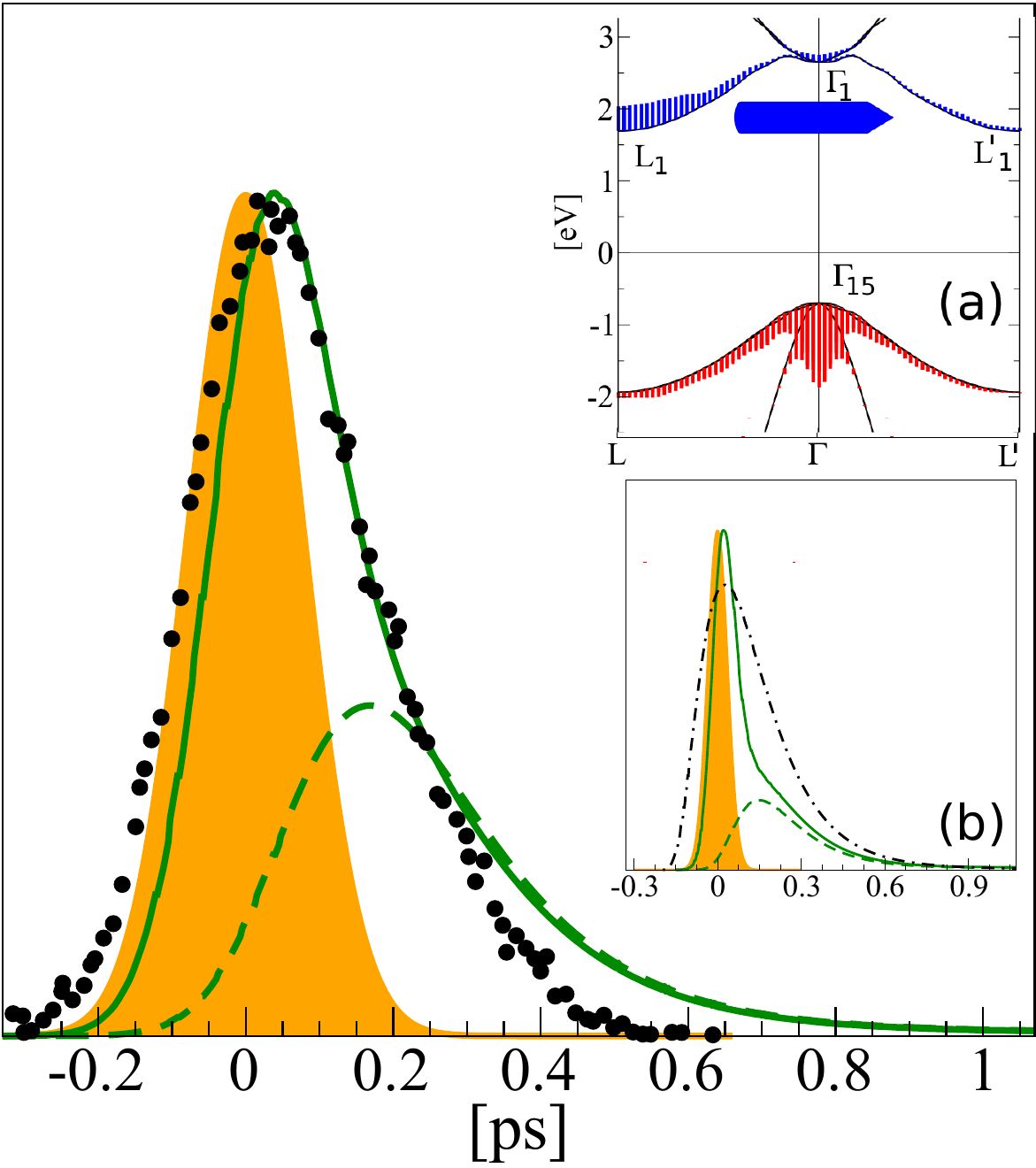}
\caption{The time dependent occupation of the $L_1$ (green continuous line) and $L'_1$
(green dashed line) levels are compared with experimental data (black dots) from \ocite{Ichibayashi2009}.
The envelope of the laser pulse is also represented (orange shadow).
In the inset $(a)$ the carriers population (electrons in blue, holes in red) is super imposed on the
band--structure at $t=0$.
The blue arrow indicates the direction of the ultra--fast $L_1\rightarrow L'_1$ scattering process.
In the inset $(b)$ the dynamics with a shorter laser pulse ($\sigma=50$\,fs) 
is compared with a {\em gedanken experiment} (black dot--dashed line) where
the same density of carriers is placed {\em by hand} at $t=0$ in the $\Gamma_{15}$ state. With the shorter pulse,
the difference between the fast $L_1\rightarrow L'_1$ scattering and the slower $L_1\rightarrow X_1$ transitions
becomes evident.
}
\label{fig:Si_2ppE}
\end{figure}

\lab{sec:Si}
Bulk Silicon is a paradigmatic material from different point of views. It is an indirect
semiconductor with moderate excitonic effects~\cite{Albrecht1998}. The fact that is indirect makes it impossible to model with simple parametric Hamiltonians,
like $H^{eq}_{model}$. At the same time it has been extensively used as a test-bed for several theories and
phenomena~\cite{Bernardi2014,Schultze2014,Shimano2011,Suzuki2009,Olevano2001,Hase2012/04//printb,Hase2003/11/06/printa}. In \ocite{Ichibayashi2009}  
a $Si$ wafer, oriented both
along the $[111]$ and the $[100]$ surface directions, is excited with a laser pulse at room temperature.
The photo--excited sample is, then, probed with a second laser pulse
that photo--emits in the continuum the excited carriers. The photo--emitted current of electrons
is measured as a function of the time delay between the pump and the probe. The final
result is a measure of the time--dependent occupation of the valence bands (represented by the dots in the main frame of Fig.\ref{fig:Si_2ppE}).
More specifically the population of carriers near the point $L_1$, i.e. at $E\approx 1.6\ eV$
above the Fermi level is probed.

The agreement between theory and experiment is excellent. Both the gradual filling  and emptying of the $L_1$
state follows quite nicely the experimental curve.
The theoretical results correctly describe the ultra--fast decay time--scale\,($\sim 180$\,fs) and
the $40\ fs$ shift of the population peak from the maximum of the pump pulse. 
The $40\ fs$ delay reflects the delicate balance
between the photo--excitation and the e--p scattering and can only be described by treating both
processes on the same footing.

Experimentally~\cite{Ichibayashi2009} the ultra--fast decay of the $L_1$ state is interpreted as due to
$L_1  \rightarrow X_1$ transitions. However a more deep analysis of the theoretical result reveals a very different scenario. 

In \figlab{fig:Si_2ppE}{a}, the population of the levels at $t=0$ is shown.
Blue lines represent charges added and red lines charges removed. The band structure is
computed along the $L \Gamma L'$  high symmetry path in the Brillouin Zone\,(BZ).
In bulk $Si$ $L$ and $L'$ are equivalent points but \fig{fig:Si_2ppE}.a shows that
the level $L'_1$ is not populated and most of the carriers are injected in the $L_1$ level.

This symmetry breaking mechanism is made possible by the external field ($\uu{V}_\kk$ operator in \e{eq:carr.1})
which, in the 2PPE experiment, is polarized along the crystallographic 
$\[111\]$ direction. This 
breaks the $L \leftrightarrow L'$ symmetry as the operation that moves $L$ in $L'$, although
being a symmetry of the unperturbed system, does not leave the $\[111\]$ direction unchanged.
In practice this means that Eq.\ref{eq:carr.1} does not respect this symmetry anymore and
$\kk$--points connected by a rotation that does not leave the pumping field unchanged are populated
in a different way.
Electrons are injected 
in the conduction band along the $\Gamma-L$ line but not, for symmetry reasons, along the $\Gamma-L'$
line.
This is clearly shown in \figlab{fig:Si_2ppE}{a} where the population of the 
$L'_1$ state is represented with a dashed line.
The $L'_1$ state is gradually filled while $L_1$ is depleted revealing that the real source of the ultra--fast
decay observed experimentally is the $L_1\rightarrow L_1'$ scattering.

The case of bulk Silicon, as introduced in \sec{sec:NEQ_lifetimes}, well represent a key aspect of the time--resolved experiments, like TR--ARPES: the
excitation process can induce processes that are not present in the equilibrium, fully interacting system, but that are induced by the driving field. Thus it
is not possible to say that, in general, the time--resolved experiments simply monitor real--time, the evolution of a naturally excited system.

\subsubsection{Photocarrier-induced band-gap renormalization and ultrafast
charge dynamics in black phosphorus}
\lab{sec:BP}
Black phosphorus\,(BP) is a layered puckered structure of P atoms with remarkable anisotropic 
optical\cite{aniso2}, electrical\cite{aniso1} and thermal properties\cite{aniso4}. In contrast to other layered semiconductors, the band gap in BP can 
be adjusted from 0.33 to 2.0 eV depending on the number of layers\cite{Tran2014}. 
\begin{figure}[H]
\centering
\includegraphics[width=0.8\textwidth]{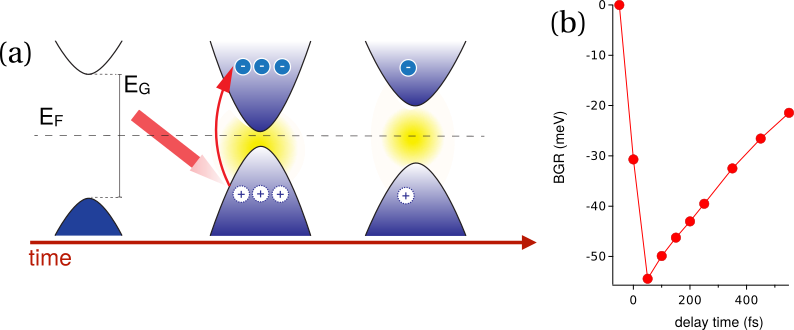}
\caption{(a) Schematics of the BGR mechanism. Before optical excitation the Fermi level\,($E_F$) lies within the band-gap and the valence band is fully occupied (left panel).
Upon optical excitation electrons are photoexcited in the conduction band and high-energy holes are left in the valence. These high-mobility charges are
responsible for the band-gap reduction (central panel). The equilibrium bang-gap energy is recovered in time (right panel). (b) Simulated evolution of the BGR as a
function of time. The maximum is reached slightly later than the maximum of the pump pulse.}
\label{fig:BP_bgr}
\end{figure}

BP is a direct gap semiconductor characterized by a wide lowest conduction band, with a large concavity that, together with the additional valleys induced by
the complex electronic structure of the material dictates most of its dynamical properties.

Indeed, the response of BP to the external optical excitation is more complex than just a variation of the occupancy of the valence and conduction states.
Namely, in \ocite{Roth2019} it has been observed a sudden change in the energy position of the valence band, following the arrival of the pump pulse. 
The simulations performed using the \yambo code have attributed this
energy shift to an ultrafast band-gap renormalization\,(BGR), as schematically illustrated in \figlab{fig:BP_bgr}{a}. The band-gap shrinks (central panel), and then
relaxes back to equilibrium on a longer timescale (right panel). 

\begin{figure}[H]
\centering
\includegraphics[width=0.8\textwidth]{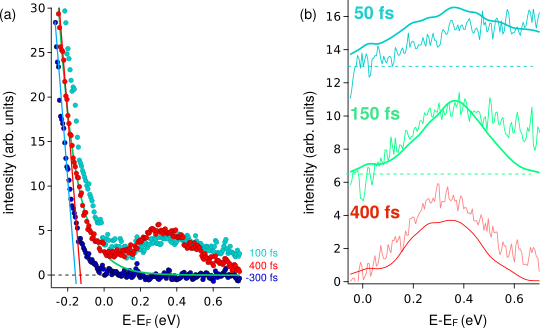}
\caption{Quantitative comparison of the experimental and theoretical energy distribution curves (EDCs) measured at the
$\overline{\Gamma}$ point. From \ocite{Roth2019}. In frame (a) the experimental EDCs are showed together with the valence band tail. In the frame\,(b) the
experimental EDC\,(thin line) is compared with the simulations\,(thick line). The agreement is excellent.}
\label{fig:BP}
\end{figure}

\fig{fig:BP} presents a quantitative comparison of the experimental and theoretical BGR. \fig{fig:BP}{a} shows three energy distribution curves (EDCs) measured at the
$\overline{\Gamma}$ point.
A large increase in the spectral intensity within the original gap region is observed 100\,fs after the arrival of the pump. \figlab{fig:BP}{b}
compares experimental EDCs at the $\overline{\Gamma}$ point  for three delay times [50 fs (teal), 150 fs (green) and 400 fs (red)], with theoretical
calculations (thin lines). The experimental data are plotted after subtracting an exponential function that fits the tail of the VB spectral intensity, as shown
by the green line in \figlab{fig:BP}{a} for the 400\,fs EDC. The excitation of hot electrons at high energy in the CB is clearly visible at 50 fs, as a broad and
structureless distribution. The simulations show that the deep CB valley around the bulk $\Gamma$ point accommodates most of the directly excited carriers.

On a longer time scale, electron--electron and electron--phonon scattering mediates the relaxation of the charges towards the CBM at the bulk $Z$
point, where they accumulate. The remarkably good agreement between  experimental and calculated line shapes at 250 fs and 400 fs indicates that
the measured ARPES peak reflects the density of states of the CB. This population evolves at larger delay times due to electron-hole recombination with the VB.
Additional experimental and theoretical details on the excitation and the following cascade process are provided in the Supporting information.

\section{Conclusion and perspectives}
\ppe experiments are nowadays performed on a wide family of materials that go from bulk solids to nanostructures, passing by atoms and molecules. At the same
time the theoretical approaches are based on different methods and approximations. The scenario that emerges is of an exciting new field where still there is
room for many  years of inspiring research.

Already in the equilibrium regime the scientific community has witnessed the gradual merging of Many--Body techniques with Material Science methods. This
process led to the birth of the \ai\, methods, which applies the well--known diagrammatic concepts in the Density--Functional Theory representation from which
it takes accuracy and predictivity. 

What is happening now is the gradual merging of Density--Functional Theory with Non--Equilibrium Many--Body techniques. As discussed in this review, however,
there are some key differences with the equilibrium case, where there exists a well established and tested approach: the $GW$/$BSE$, reviewed in several works
and routinely applied  as coded in several public scientific codes.

In the Non--Equilibrium Many--Body approach, at the moment, there are very few scientific codes available that, in addition, use different approaches whose accuracy is still
not totally assessed. Indeed the test\&validate process that has been used to establish the accuracy of the  $GW$/$BSE$ approach requires many years and a large
community of users. At the same time also the very basic physical processes that are triggered by the driving pump field are still the subject of an intense
debate. 

The excitonic concept is a striking example of the current state-of--the--art. Considering the linear--response regime, where a tiny fraction of carriers are
excited in the empty states, physical intuition would suggest that the photo--excited electrons and holes would bound together in an excitonic state. 
Is this what really the Non--Equilibrium Many--Body approach predicts to happen? And in case this is the right physics how can be experimentally observed?
Moreover, what happens when the realistic material is considered with all relaxation processes considered?

At the moment we do not have a definitive answer to this question. But currently there is a very active research in the field that can  count on the unprecedented collaboration 
of theory, models and \ai\, methods and codes, like the {\it Yambo} code. In addition we are gradually creating bridges with the experiments in order to
unambiguously interpret the measured quantities, like the TR--ARPES, in terms of microscopic processes. 

In the future we plan to continue interacting with experiments in order to assess the numerical tools based on the  Non--Equilibrium Many--Body technique. We
are also working to derive new theoretical schemes to be implemented in the {\it Yambo} code after being tested on simple models. The scenario clearly points to
the need of an unprecedented collaboration of experimentalists, theoreticians and material scientists to create an universal set of public tools to be used in the
scientific community  providing an accurate guide  for the  interpretation of time--resolved experiments, like the  TR--ARPES.

\section{Acknowledgments}
A.M. acknowledges the funding received from the European Union projects: MaX {\em Materials design at the eXascale} H2020-EINFRA-2015-1, Grant agreement n.
676598, and H2020-INFRAEDI-2018-2020/H2020-INFRAEDI-2018-1, Grant agreement n. 824143;  {\em Nanoscience Foundries and Fine Analysis - Europe} 
H2020-INFRAIA-2014-2015, Grant agreement n. 654360. G.S. and E.P. acknowledge the financial support from MIUR PRIN (Grant 
No. 20173B72NB), from INFN through the TIME2QUEST project, and from  Tor Vergata University through the Beyond Borders Project  ULEXIEX.

\section*{Bibliography}

\end{document}